\documentclass[prl,twocolumn,superscriptaddress]{revtex4-2}
\usepackage{amsmath,amssymb,mathrsfs}
\usepackage{graphicx}
\usepackage{colortbl}
\usepackage{braket}
\usepackage{CJK}
\usepackage{bm}
\usepackage{amsfonts}
\usepackage{ulem}
\begin{document}
\title{Exact Fisher zeros and thermofield dynamics across a quantum critical point}

\author{Yang Liu}
\affiliation{Key Laboratory of Polar Materials and Devices (MOE), School of Physics and Electronic Science, East China Normal University, Shanghai 200241, China}

\author{Songtai Lv}
\affiliation{Key Laboratory of Polar Materials and Devices (MOE), School of Physics and Electronic Science, East China Normal University, Shanghai 200241, China}

\author{Yuchen Meng}
\affiliation{Key Laboratory of Polar Materials and Devices (MOE), School of Physics and Electronic Science, East China Normal University, Shanghai 200241, China}

\author{Zefan Tan}
\affiliation{Key Laboratory of Polar Materials and Devices (MOE), School of Physics and Electronic Science, East China Normal University, Shanghai 200241, China}

\author{Erhai Zhao}
\altaffiliation{ezhao2@gmu.edu}
\affiliation{Department of Physics and Astronomy, George Mason University, Fairfax, Virginia 22030, USA}

\author{Haiyuan Zou}
\altaffiliation{hyzou@phy.ecnu.edu.cn}
\affiliation{Key Laboratory of Polar Materials and Devices (MOE), School of Physics and Electronic Science, East China Normal University, Shanghai 200241, China}

\begin{abstract}
By setting the inverse temperature $\beta$ loose to occupy the complex plane, Fisher showed that 
the zeros of the complex partition function $Z$, if approaching the real $\beta$ axis, reveal a thermodynamic phase transition. 
More recently, Fisher zeros were used to mark the dynamical phase transition in quench dynamics.
It remains unclear, however, how Fisher zeros can be employed to better understand quantum phase transitions
or the non-unitary dynamics of open quantum systems. 
Here we answer this question by a comprehensive analysis of  
the analytically continued one-dimensional transverse field Ising model. 
We exhaust all the Fisher zeros to show that in the thermodynamic limit they congregate into a remarkably simple pattern 
in the form of continuous open or closed lines. These Fisher lines evolve smoothly as the coupling constant is tuned,
and a qualitative change identifies the quantum critical point. 
By exploiting the connection between 
$Z$ and the thermofield double states, we obtain analytical expressions for the short- and long-time dynamics of the
survival amplitude, including its scaling behavior at the quantum critical point.
We point out $Z$ can be realized and probed in monitored quantum circuits.
The exact analytical results are corroborated by the numerical tensor renormalization group. We further show that similar patterns of Fisher zeros also emerge in other spin models.
Therefore, the approach 
outlined may serve as a powerful tool for interacting quantum systems.

\end{abstract}

\maketitle
{\it Introduction. }
There is a renewed interest in the concept of complex-valued partition function and its zeros
in quantum many-body physics.
In the seminal work of Lee and Yang
 on phase transitions, the magnetic field in spin Hamiltonians 
 (or the chemical potential in a grand canonical ensemble) is analytically continued to 
 take on complex values~\cite{LeeYang1,LeeYang2}. Then phase transitions in the thermodynamic limit
 can be inferred and analyzed by tracking the zeros of the complex-valued partition function $Z$
 evaluated for finite-size systems. Fisher extended the recipe by
 allowing the inverse temperature $\beta=1/k_BT$ 
to be complex instead, which applies to all systems~\cite{fisher1965statistical}. 
Yet, compared to the Lee-Yang zeros, the Fisher zeros are more 
challenging to find or analyze. 
Because $\beta$ couples to every term in the Hamiltonian,  
the locations of Fisher zeros on the complex $\beta$ plane
appear to feature less regularity. They may not form simple smooth curves
as the thermodynamic limit is approached~\cite{bena2005statistical}, 
 in sharp contrast to the celebrated circle theorem for Lee-Yang zeros~\cite{LeeYang1,LeeYang2,Flindt2021prr}. To unveil the unifying patterns behind the ostensible complexity of Fisher zeros,
exact analytical results would be greatly helpful.

\begin{figure}[t]
  	\centering
  	\includegraphics[width=0.47\textwidth]{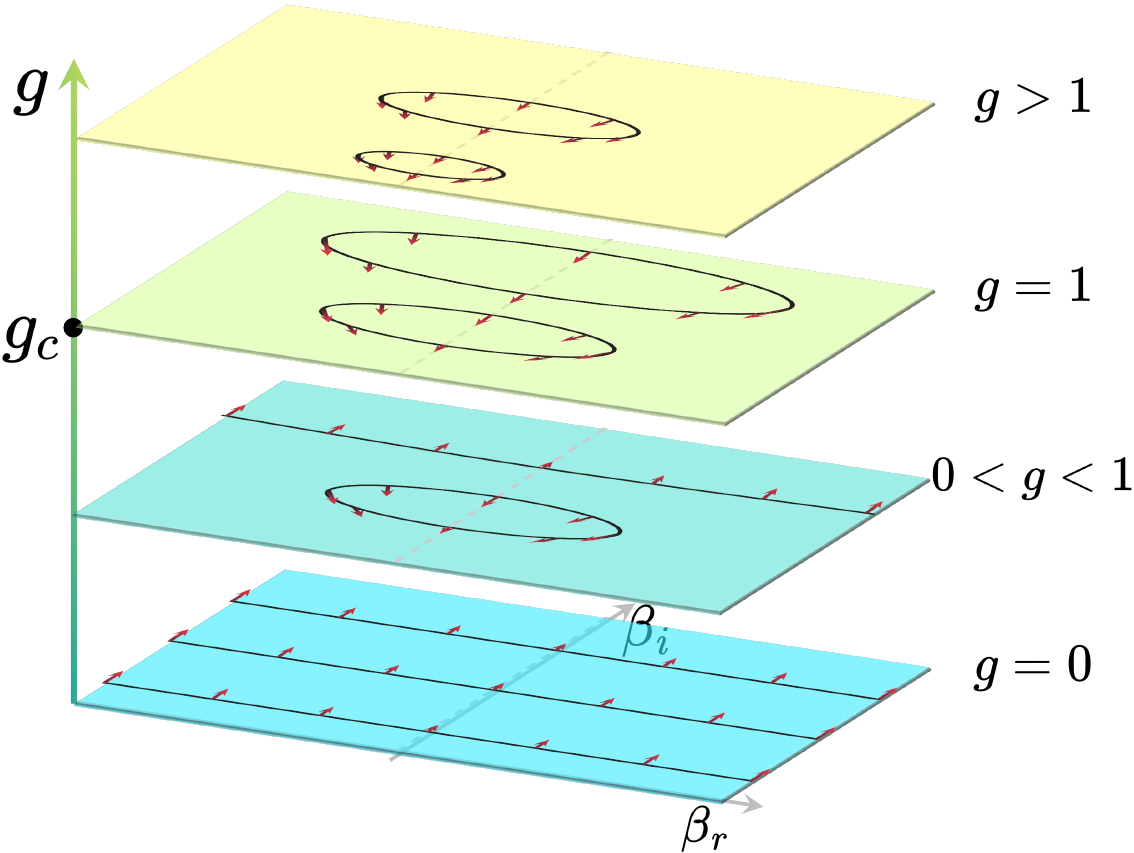}
  	\caption{The continuous lines of Fisher zeros on the complex $\beta=\beta_r+i\beta_i$ plane for 1DTFIM (schematic). The arrows indicate the velocities of the zeros as the transverse field $g$ is increased. At $g=0$, the zeros congregate into a set of open lines. As $g$ increases, closed lines (ovals) of zeros appear, while the open lines shift upwards to larger $\beta_r$ and eventually vanish at the QCP ($g=g_c=1$). The closed lines continue to move toward the origin as $g$ is further increased beyond $g_c$.}
  	\label{fig:fig1}
  \end{figure} 

Fisher zeros were invented to comprehend thermodynamic phase transitions.
Recently, several of us~\cite{liu2023CPL} proposed to use them to
 probe quantum phase transitions~\cite{Sachdevbook}. Numerical data on the
 one-dimensional transverse field Ising model (1DTFIM) reveal that 
 as the coupling constant (the transverse field) $g$ is tuned,
 the motion of certain Fisher zeros 
correlates with the energy gap associated with the domain wall excitations.
The analysis of Ref.~\cite{liu2023CPL} was limited to $g\leq 1$,
so a natural question is what happens on the disordered side $g>1$.
Naively, one might expect from the Kramers-Wannier (KW) duality~\cite{KW1941}
that the zeros at $g>1$ simply mirror those at $g<1$. This conjecture turns out to be 
false. We shall show that the KW duality breaks down for complex $\beta$. To establish a
correct, complete picture of Fisher zeros across the quantum critical point (QCP), an exact formula of $Z$ valid for all $g$ values is required to delineate the intricate interplay of quantum and thermal fluctuations in various regions of the complex $\beta$ plane. 

An even more significant impetus for complex $Z$ comes from the flurry of
experiments on quantum simulators and quantum circuits which prompted the generalization of
the notion of phase transitions to time-dependent and open 
quantum systems~\cite{fisher2023random}. The complex zeros of the Loschmidt echo amplitude were used to define 
dynamical phase transition in 
quench dynamics~\cite{DynamicalPT2013PRL}. 
Following the discovery of measurement induced phase transitions in
nonunitary quantum circuits~
\cite{PhysRevB.98.205136,PhysRevX.9.031009,PhysRevB.99.224307},
various non-Hermitian generalizations of the Ising model have been considered 
where the coupling constants become complex valued 
as a result of the non-unitary evolution introduced by measurements. 
For example, Ref.~\cite{PhysRevResearch.4.013018} mapped a non-unitary quantum circuit to a
2D classical Ising model at complex temperature.
These developments hint a new era of complex $Z$ in which the complexification
is no longer just a mathematical trick but mandated by experiments to take on physical 
significance. 

This work is  
inspired by the confluence of these ideas.
By exhausting the Fisher zeros for a canonical quantum many-body system, the 1DTFIM,   
 we describe how insights about quantum phase transitions and
quantum dynamics can be gleaned from the complex-valued 
 partition functions ${Z}(\beta)$. A few surprising results are obtained. (i) In the thermodynamic limit, 
 the Fisher zeros of 1DTFIM form continuous curves (lines and loops) which 
move smoothly as the coupling constant $g$ is tuned.
(ii) Even though the Fisher zeros never touch the real $\beta$ axis, the quantum critical point can be 
unambiguously inferred from either the vanishing of open lines (Fig.~\ref{fig:fig1}) or 
the scaling of low-energy excitations extracted from ${Z}$. 
(iii) We exploit the connection between ${Z}$ and the maximally entangled 
thermofield double (TFD) states~\cite{takahashi1996thermo}, which are dual to eternal black holes 
according to the AdS/CFT correspondence~\cite{Maldacena:2003aa},
to make analytical predictions about the dynamics of TFD. These include the short-time
decay of the survival probability, 
its periodic oscillation at low temperatures, and the recurrence time at the quantum critical point. 
(iv) We further show that 
aside from being a useful theoretical device, ${Z}$ can be realized and
probed in nonunitary quantum circuits where the unitary evolution of the qubits is interspersed with
projective measurements and post selection.

{\it Fisher zeros from the exact partition function}.
The Hamiltonian of the 1DTFIM can be written as
\begin{equation}
   {H} = -\sum_{i=1}^{L} \left(\sigma_i^z\sigma_{i+1}^z + g\sigma_i^x\right),
    \label{eq:Hamtransf}
\end{equation}
where $\sigma^x_i$ and $\sigma^z_i$ are the Pauli matrix operators on site $i$,
and the coupling constant $g$ is the transverse magnetic field measured in units of
the nearest neighbor Ising interaction (set to one). As is well known, in the thermodynamic limit ($L\rightarrow\infty$),
a quantum phase transition at the critical coupling $g_c = 1$ separates 
the ferromagnetic phase ($g<1$) from the paramagnetic phase ($g>1$).

While 1DTFIM is exactly solvable, the analytical form of its partition function $Z$ is subtle to write down.
Usually the model is treated in fermionic representation after the Jordan-Wigner transformation, 
where one has to confront the two sectors of the total fermion number~\cite{NiveditaPRE2020}.
We find it cleaner to stick to the spin representation and map $Z=\mathrm{Tr}e^{-\beta H}$ to
that of an anisotropic 2D Ising model and its Onsager-Kaufman solution~\cite{Onsager1944,Kaufman1949}, 
by Trotter decomposition along the imaginary time axis $\beta$ which plays the role of another 
spatial dimension $y$~\cite{Suzuki1976}. The result is
\begin{align}
	Z = &\frac{1}{2}
	\left[
	\prod_{k=1}^L 2\cosh\left(\beta\epsilon_{2k}\right)+\mathrm{sign} (g_c-g)\prod_{k=1}^L 2\sinh(\beta\epsilon_{2k})\right.
	\nonumber
	\\
	&\left.+\prod_{k=1}^L 2\cosh(\beta\epsilon_{2k-1})+\prod_{k=1}^L 2\sinh(\beta\epsilon_{2k-1})
	\right],
	\label{eq:suzuki}
\end{align}
where 
$\epsilon_k = [1 + g^2 - 2g\cos(\pi k/L)]^{1/2}$.
We refer to  Eq. \eqref{eq:suzuki} as the Suzuki solution, even though the all important 
sign$(g_c-g)$ did not appear explicitly in Ref.~\cite{Suzuki1976} and perhaps is not widely appreciated. 
Without the sign factor, one would have expected the KW duality~\cite{KW1941}, i.e.
$Z$ is invariant under the transformation $g\rightarrow 1/g$, $\beta\rightarrow g\beta$ so that
it is sufficient to analyze the subspace $g<1$.
This is only an illusion: the correct 
formula Eq.~\eqref{eq:suzuki} shows that $Z$ itself does not possess KW duality and implies the presence of the recently proposed non-invertible symmetry~\cite{Seiberg2024KW,Senthil2024}. 
The distribution of Fisher zeros on the quantum disordered side $g>1$ is qualitatively different from
the ordered side $g<1$. 
From the expression of $Z$, Fisher zeros can be located and visualized by following the procedure outlined in \cite{supp}.

{\it Tensor renormalization group with complex $\beta$}. We also compute $Z$ using an independent numerical method
which serves two purposes. First, it confirms the analytical solution Eq.~\eqref{eq:suzuki} is correct. 
Second, it demonstrates a general and precise technique to compute $Z(\beta\in \mathbb{C})$ for interacting 
quantum spin models for which there is no exact solution. 
Since $Z$ can be represented as the trace of a product of local tensors, it can be evaluated by tensor network 
algorithms~\cite{ORUS2014117,TNreview1,TNreview2} which have proven to be efficient and accurate for spin models~\cite{Zou2014PRD}. 
Starting from the 1D system, 
we first construct a 2D ``lattice" of size $L\times N$ by Trotter decomposition $\beta=\tau N$. Then, $Z$ is obtained from the high order tensor renormalization group (HOTRG)~\cite{XieHOTRG}. We have checked that tensor results are in agreement with both exact diagonalization and the exact solution above (see~\cite{supp} for details).

It is worth to mention that when $\beta$ is complex, 
Monte Carlo develops sign problems but HOTRG can yield $Z$ accurately~\cite{Zou2014PRD}. 
Previously, renormalization group (RG) 
transformation has been generalized to the complex $\beta$ plane to
study confinement in lattice gauge theory and classical $O(N)$ models. It was observed that the Fisher zeros are 
located at the boundary of the attraction
basins of infrared fixed points, and they control the global properties of the complex RG flows~\cite{RGflow2010,Zou2011PRD}.
Here we 
confine our attention to the evaluation of Fisher zeros through HOTRG
but for quantum spin models.

{\it Motion of Fisher zeros across QCP}. 
The exact solution corroborated by HOTRG yields
a complete picture, sketched in Fig.~\ref{fig:fig1}, of 
the evolution of Fisher zeros on the complex $\beta$ plane as the coupling constant $g$ is varied. 
The main observations are summarized into (i) to (iv) below. (i) The Fisher zeros of 1DTFIM 
appear as continuous lines, rather than isolated points or densely packed regions, in the thermodynamic limit
$L\rightarrow \infty$.
(ii) For $g<1$, there exist {\it open lines} of zeros that are approximately parallel to the $\beta_r$ axis.
(iii) As $g$ increases from zero to one, these open lines shift upwards toward large $\beta_i$,  
and vanish completely at the QCP $g=1$.
(iv) Away from $g =0$, {\it closed loops} of Fisher zeros also appear and persist for all $g$. 
These loops are centered around the $\beta_i$ axis and move toward $\beta_i=0$ as $g$ is increased.
We stress that properties (i) to (iv) came as surprises, in the sense that they are not obvious 
from staring at the model or the exact solution of $Z$.

Some of these features can be appreciated by considering a few limit cases. 
At $g=0$, $Z=\mathrm{Tr} V^L$ where the transfer matrix $V=e^\beta + \sigma_x e^{-\beta}$.
The Fisher zeros are determined by the condition 
$\tanh \beta=e^{i n\pi/L}$ with $n$ an integer, i.e., they are uniformly distributed on the unit
circle on the plane of $\tanh\beta$~\cite{PhysRevResearch.4.013018}. On the complex $\beta$ plane, the zeros form a set of straight lines 
at $\beta_i=\pi/4+m\pi/2$, with $m$ an integer,  as $L\rightarrow \infty$.
Another simple case is the small size system with $L=2$. After some algebra, one finds the zeros are located 
on the $\beta_i$ axis at 
\[
\beta_i = (m+ \frac{1}{2})\pi/(\sqrt{1+g^2}\pm 1)
\]
where $m$ again is an integer. While the size $L=2$ is too small to feature Fisher zeros away from the  $\beta_i$ axis
or the closed loops, it does correctly capture the trend that the zeros move toward $\beta_i=0$ as $g$ is increased.
We note that previous work~\cite{Yin2024prb} focused on zeros on the imaginary $\beta$ axis originated from the $\cosh$ terms in $Z$.

It is clear from (iii) that the QCP 
is marked by a qualitative change in 
the Fisher zeros. To describe the lowest open Fisher line, one can take its imaginary part $\beta_i$
at some fixed large value of $\beta_r$ and plot $1/\beta_i$ against $g$. In Ref.~\cite{liu2023CPL}, it was shown
that the resulting curve is linear near QCP. 
Similarly, for the closed Fisher lines, we can select a characteristic point 
on each curve, 
e.g., recording the $\beta_i$ value of the rightmost point. 
Then a linear relationship between $1/\beta_i$ and 
$g$ is observed on the quantum disordered side $g>1$, which is consistent with the spin flip excitations.
A more refined description is to define
the velocities of the zeros  
on the complex $\beta$ plane with increasing $g$. 
Interestingly, we find the speed of the zeros exhibits properties similar to the Gr\"uneisen parameter~\cite{Zhang2019prl,Zhang2023cpl}, see \cite{supp}.

\begin{figure}[h]
    \includegraphics[width=0.46\textwidth]{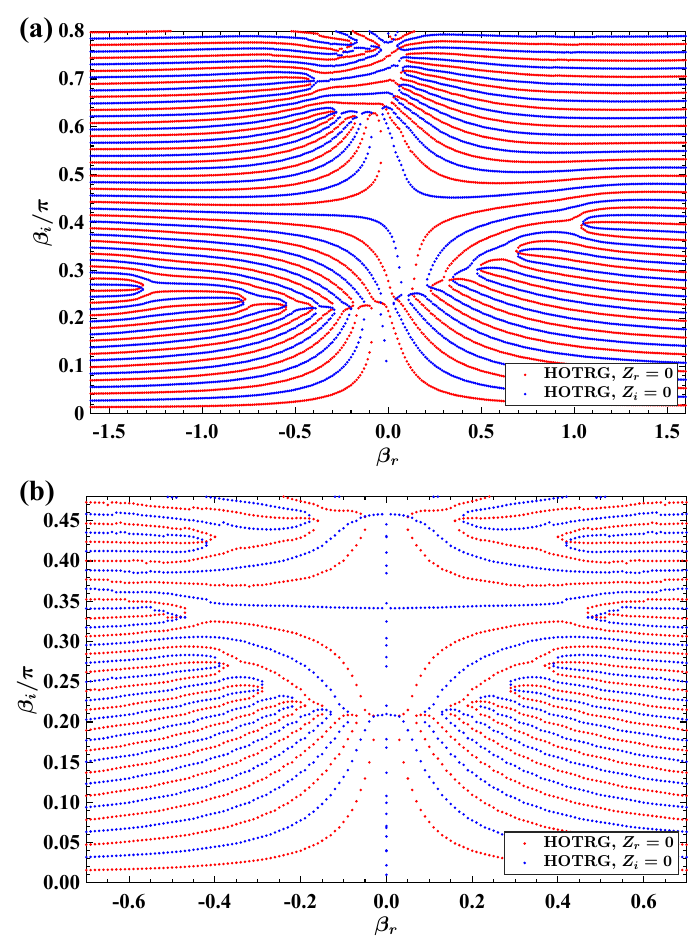}
    \caption{The pattern of Fisher zeros, as the intersections of $Z_r=0$ (red) and $Z_i=0$ (blue) computed by HOTRG for finite-size systems ($L=32$), on the complex $\beta$ plane. (a) The zeros for
the ANNNI model with $\kappa=0.1$ and $g=0.5$, calculated with D-bond $D_b=30$, form an open line near the 
$\beta_r$-axis. (b) In the XY model with $\gamma=0.01$, calculated with $D_b=50$, the zeros near the $\beta_r$-axis form closed loops.
  \label{fig:fig2}}
\end{figure}

{\it Fisher zeros in other spin models.}
We conjecture that the regularity of Fisher's zeros   
and its correlation with QCP, observed above for 1DTFIM, 
generalize to a host of other systems. 
Since ${Z}$ and its zeros can be reliably computed from the tensor network method,
the program outlined here can serve as a powerful diagnostic tool for quantum many-body physics.
To illustrate this, we consider two examples. 
The first is the axial next nearest neighbor Ising (ANNNI) model with the Hamiltonian
\[
H=-\sum_{i=1}^{L} \left(\sigma_i^z\sigma_{i+1}^z -\kappa\sigma_i^z\sigma_{i+2}^z+g\sigma_i^x\right),
\]
where we choose $\kappa=0.1$ and $g=0.5$ so the system has a gap. 
The second is the $XY$ model with the Hamiltonian
\[
H=\sum_{i=1}^{L}[(1+\gamma)\sigma_i^x\sigma_{i+1}^x+(1-\gamma)\sigma_i^y\sigma_{i+1}^y],
\]  
where we chose $\gamma=0.01$, bringing the system close to a 
QCP. For both systems, numerical calculations 
show that as $L$ increases, the zeros indeed approach continuous lines. Figure~\ref{fig:fig2} 
compares the Fisher zeros of these two models obtained from HOTRG with $L=32$~\cite{supp}. 
For the ANNNI model, open lines corresponding to domain wall excitations still exist. But the zero locations are no longer symmetric between $\beta_r>0$ and $\beta_r<0$, causing 
the open lines to tilt. 
For the XY model near QCP, we witness a similar qualitative change as seen in the 1DTFIM: 
the open lines completely vanish, giving way to closed loops.
The change is mandatory because otherwise, the two phases separated by the QCP would share the same global patterns of RG flows on the complex $\beta$ plane, which would contradict the fact that
they correspond to different RG fixed points.

{\it What can $Z$ tell us about quantum dynamics}? 
We now view
$Z$ from another perspective, as the survival amplitude of a certain quantum state. 
Let $\beta=\beta_r+it$, and interpret the imaginary (real) part of $\beta$ as time $t$ 
(the inverse temperature $\beta_r=1/k_BT$). Let $E_n$ be the eigen energies, and $|n\rangle$ the corresponding eigen states,
of Hamiltonian $H$. Then   
\[
Z(\beta_r,t)=\sum_n e^{-\beta_r E_n}e^{ -iE_n t}
\]
describes the unitary dynamics of a mixed ensemble.
The modulus square of $Z$ defines the spectral form factor 
\[
S(\beta_r,t)=\left|\frac{Z(\beta_r,t)}{Z(\beta_r,0)}\right|^2,
\]
which is normalized to 1 as $t=0$. The disorder average of $S$
has been used to diagnose chaos and information scrambling in Hamiltonians containing random couplings, e.g.,
the SYK model \cite{cotler2017black}. Following Ref.~\cite{PhysRevD.95.126008}, we  
purify the mixed state above by introducing the thermofield double state in an enlarged Hilbert space
\[
|\psi(\beta_r,0)\rangle = \frac{1}{\sqrt{Z(\beta_r,0)}}\sum_n e^{-\beta_r E_n/2}|n\rangle_L\otimes|n\rangle_R,
\]
where, for clarity, the subscripts $L,R$ are used to denote the left and right copy, respectively~\cite{takahashi1996thermo}.
Under the single Hamiltonian $H_L\otimes 1_R$,
the TFD state evolves with time as 
$\psi(\beta_r,t)= e^{-it(H_L\otimes 1_R)}|\psi(\beta_r,0)\rangle $.
One finds that $S$ is nothing but the survival probability~\cite{PhysRevD.95.126008}
\[
S(\beta_r,t)=|\langle \psi(\beta_r, t)|\psi(\beta_r,0)\rangle |^2.
\]
In contrast to the Loschmidt echo in quench dynamics~\cite{DynamicalPT2013PRL},
here the connection between $Z$ and
quantum dynamics is established by the TFD states, which are dual to the eternal black hole
in the context of AdS/CFT correspondence~\cite{Maldacena:2003aa}.
Via this bridge, the detailed knowledge about the analytical
structure of $Z$, including the Fisher zeros, can now be translated into new insights about quantum dynamics.
Three examples for the 1DTFIM are given below.

\begin{figure}[t]
    \centering
    \includegraphics[width=0.45\textwidth]{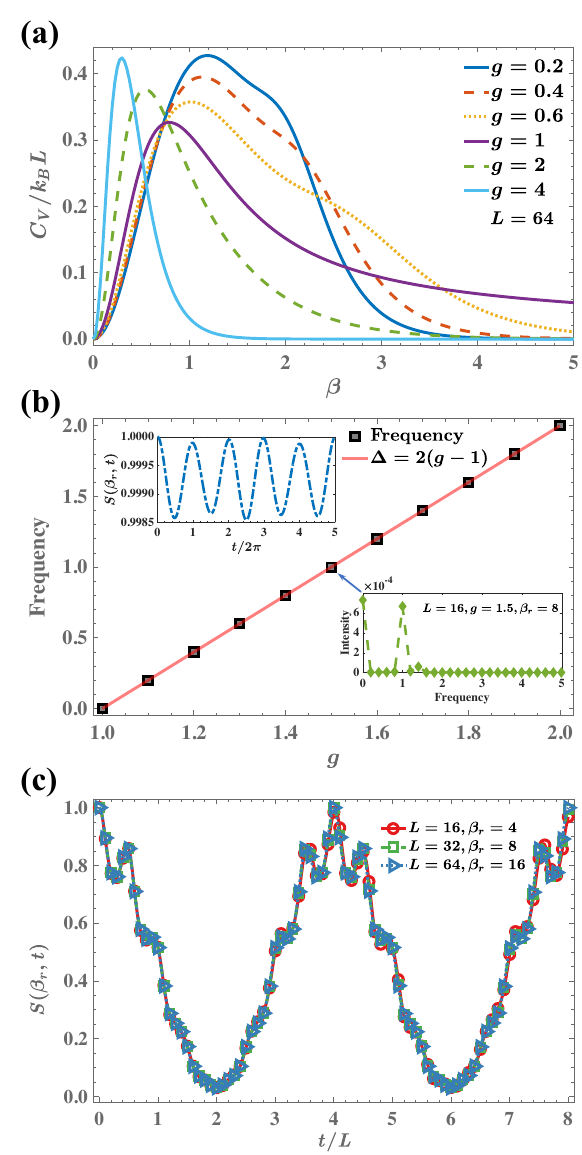}
    \caption{The rich behaviors of the spectral form factor (survival probability) $S(\beta_r,t)$. (a) 
    The short-time decay of $S$ is dictated by the specific heat capacity $C_V/L$. (b) The low temperature oscillations
    (inset, $g=1.5$) of $S$. The oscillation frequency obtained from Fourier transform can be fit by $\Delta= 2(g-1)$
    and vanishes at QCP. (c) Fine structures of $S$ over long time scales for  
    different $\beta_r$ and $L$. At the QCP, all exhibit the same periodicity $t^*=4L$. The overlay of curves shows self-similarity, $S(\beta_r,t)_L\approx S(n\beta_r,nt)_{nL}$ with $n=2$.}
    \label{fig:fig3}
\end{figure}

The short time dynamics of $S$ is characterized by exponential decay $S(t)\simeq e^{-(t/\tau)^2}$ 
with $\tau^2 = {k_B\beta_r/C_V}$. Figure~\ref{fig:fig3}(a) shows the specific heat capacity $C_V/L$
obtained by fitting $S(t)$ at some typical values of $g$. 
The decay rate $1/\tau\propto \sqrt{C_V}$ correlates with the locations of Fisher zeros at which $S$ is forced
to vanish. This can be seen on the low $\beta_r$ portion of Fig.~\ref{fig:fig3}(a): 
as $g$ is increased from $g=1$, the Fisher zeros move closer to the $\beta_r$ axis,
and accordingly $S$ is suppressed from one to zero at an increasing rate. 
At high $\beta_r$ (low temperatures), 
however, the decay is most rapid at the QCP, where quantum fluctuation is strong and $C_V$ reaches maximum.

The long time behavior of $S(t)$ is complicated but simplicity emerges 
at low temperatures. To understand this, we can convert the products of hyperbolic sines and cosines
in the exact solution of $Z$ into sums and expand them for large $\beta_r$.
The leading contributions to $S(t)$ are
\[
S(t)\simeq 1+ 2e^{-\beta_r \Delta} \cos(\Delta t) + ...,
\]
where the oscillation frequency 
$\Delta=\sum_k(\epsilon_{2k-1}-\epsilon_{2k})$ has the meaning of the excitation gap.
Previously in Ref.~\cite{liu2023CPL}, it was noted from numerics  
that for $g\leq 1$ the oscillation frequency vanishes
at the QCP while the oscillation amplitude reaches maximum. The analytic asymptote
derived here not only explains this observation but also expands its validity to the disordered side $g\geq 1$ [Fig.~\ref{fig:fig3}(b)].
Thus, the low temperature oscillations of $Z(\beta_r,t)$ 
provide another diagnostic tool for QCP.

The dynamics of $S$ exhibits rich scaling behaviors at the QCP. For a system of finite size $L$,
the energy spectrum is discrete and the initial TFD state will come back to itself after
the quantum recurrence time $t^*$. 
To find $t^*$, take the limit of large $\beta_r$, expand the exact Suzuki solution,
and carry out the $k$-sum, we find at $g=1$  the oscillation frequency is 
$
\Delta = 2\tan({\pi}/{4L})
$
which becomes ${\pi}/{2L}$ in the limit of large $L$ to yield $t^*=4L$.
For smaller $\beta_r$, $S(t)$ deviates significantly from the sinusoidal form
to acquire fine structures, but
$t^*$ stays approximately the same~\cite{supp}.
Our prediction $t^*=4L$ agrees with the conformal field theory result from Ref.~\cite{NiveditaPRE2020} (see~\cite{supp} for details).
We also observe that $S$ exhibits self-similarity, i.e., $S(\beta_r,t)_L\approx S(n\beta_r,nt)_{nL}$ with $n$ an integer, at the QCP and for intermediate to large $(\beta_r,t)$. Figure.~\ref{fig:fig3}(c) shows the example of $n=2$.
This is consistent with the notion of criticality and provides a nice way to define RG flow in the complex plane (see \cite{supp} for details).

{\it Probing $Z$ from monitored quantum circuits}. 
Lastly, we show that $Z$ is not just a formal theoretical construct,
it can be realized and probed in quantum circuits.  
Consider an array of qubits, $j=1,2,..., L$. The standard 
single- and two-qubit Pauli gates implement unitary evolution such as 
$U_{zz}(t)=\exp ({it \sum_j \sigma^z_j\sigma^z_{j+1}})$ and $U_x(t) = \exp ({itg\sum_j \sigma^x_j })$.
Let us define the cycle operator $U_F(t)=U_{zz}(t)U_x(t)$, then repeated application of $U_F$ produces 
the time evolution corresponding to the 1DTFIM Hamiltonian, albeit in discretized form
$U(t)=[U_F(t/N)]^N$. Now let us introduce nonunitary gates in the form of 
$U_x(i\beta_r)$ and $U_{zz}(i\beta_r)$, i.e., by replacing $t$ with the imaginary time $i\beta_r$.
These gates can be realized by including ancilla qubits and performing projective measurements
and post-selection on the ancilla~\cite{PhysRevLett.130.230401,PhysRevLett.130.230402,PhysRevResearch.6.013131}, where 
the value of $\beta_r$ characterizes the strength of the measurement.
If we intersperse these unitary and nonunitary gates~\cite{fisher2023random}, e.g.,
by forming a cycle 
\[
U_F(\beta_r,t) = U_x(i\beta_r)U_x(t)U_{zz}(i\beta_r)U_{zz}(t),
\]
the resulting evolution operator yields exactly $Z(\beta_r,t)=[U_F(\beta_r/N,t/N)]^N$ in the limit
of large $N$.
From this perspective, $Z(\beta_r,t)$ describes the competition between unitary time evolution
and quantum measurement in an open quantum many-body system.

The circuit construction here differs from the approach in 
Ref.~\cite{PhysRevResearch.4.013018} based on the transfer matrix. From an alternative perspective, the nonunitary gates 
introduce imaginary parts to the coupling constants in the Hamiltonian and render it
non-Hermitian~\cite{nonHermitian1}. Previous studies on non-Hermitian Ising-type models mostly
focused on their entanglement or spectral properties~\cite{PhysRevLett.130.230401,PhysRevLett.130.230402,PhysRevResearch.6.013131}. 
Given the recent success in observing
Lee-Yang zeros~\cite{zero_exp1} and Loschmidt ratios~\cite{DQPT_exp}, it seems promising that the circuit connection to 
$Z$
and TFD states~\cite{zhu2020generation} can lead to future experimental studies of Fisher zeros~\cite{Wei2014}.

{\it Conclusion. }In summary, this work presents a comprehensive analysis of the Fisher zeros in the 1DTFIM. We have demonstrated that Fisher zeros form continuous, smoothly shifting curves in the complex $\beta$ plane, and they offer insights into quantum criticality despite never touching the real $\beta$ axis. 
The complex partition function $Z$ provides valuable analytical predictions about the quantum dynamics which is beyond the reach of the traditional thermodynamics, 
and can be realized and probed in nonunitary quantum circuits.
Our improvement/clarification over the original Suzuki solution correctly captures the different excitations on the two sides of the QPT
and the breakdown of the Kramers-Wannier duality. 
Successful generalization of the numerical analysis to other (the ANNNI and XY) spin models demonstrates that this approach can open up a promising avenue to study quantum critical systems in 1D~\cite{Wen2013,Zou2019PRL,Zou2020PRB}  
and higher dimensions~\cite{Zhu2013PRL,Liu2022,Zou2023}.

{\it Acknowledgments. }
We thank Youjin Deng, Shijie Hu, Ian McCulloch, Yannick Meurice, Tao Xiang, and Jiahao Yang for helpful discussions. This work is supported by the National Natural Science Foundation of China (Grant No.~12274126). E.Z. acknowledges the support from NSF Grant PHY-206419, and AFOSR Grant FA9550-23-1-0598.

%


\pagebreak

\newpage

\widetext
\begin{center}
\textbf{\large Supplemental Materials for ``Exact Fisher zeros and thermofield dynamics across a quantum critical point"}
\end{center}

\setcounter{equation}{0}
\setcounter{figure}{0}
\setcounter{table}{0}
\makeatletter
\renewcommand{\thefigure}{S\arabic{figure}}
\renewcommand{\thetable}{S\arabic{table}}
\renewcommand{\theequation}{S\arabic{equation}}
\renewcommand{\bibnumfmt}[1]{[S#1]}
\renewcommand{\citenumfont}[1]{S#1}
\makeatother

\section{Discussion on the Suzuki solution}

\subsection{Comparing the original Suzuki solution to HOTRG and ED}
\begin{figure}[t]
	\includegraphics[width=1.0\textwidth]{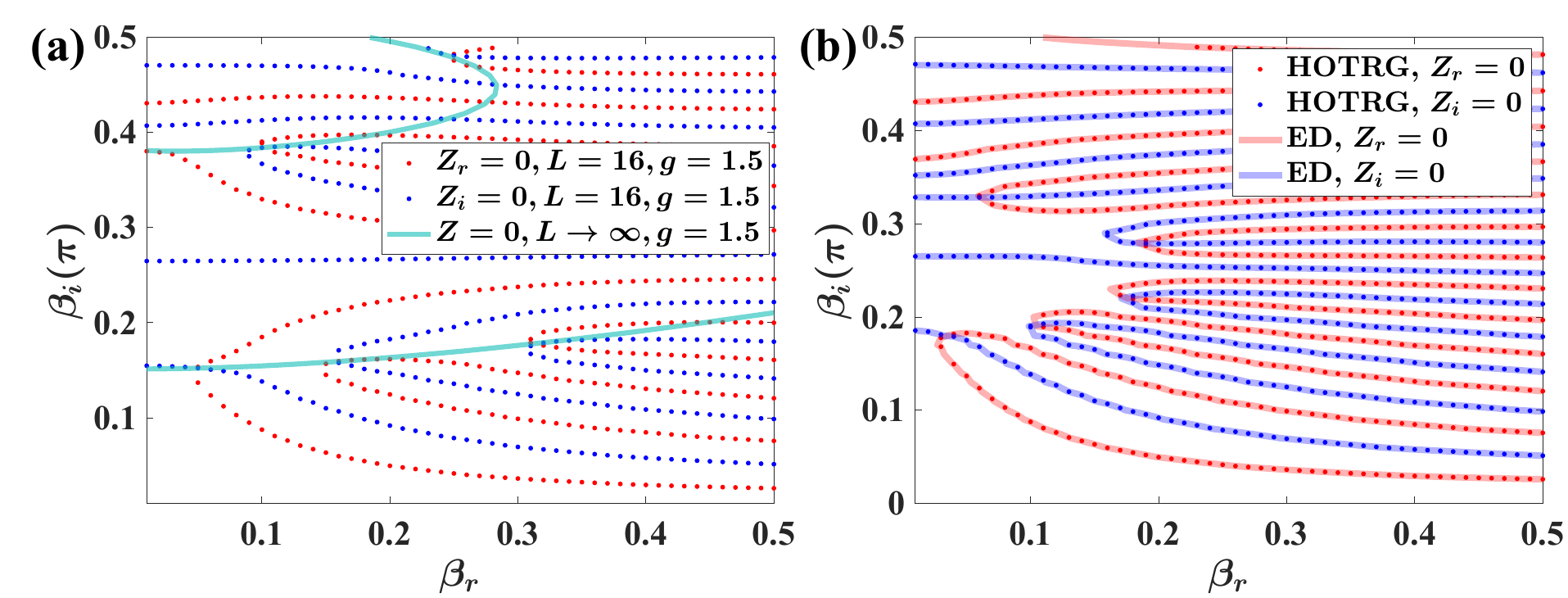}
	\caption{Fisher zero results obtained from three methods at $g=1.5$. (a) Fisher zeros for the system with $L=16$ and in the thermodynamic limit obtained using the Suzuki solution. The red and blue dots represent the values where the real ($Z_r$) and imaginary ($Z_i$) parts of the partition function become zero for $L=16$, respectively. Their intersections yield the Fisher zeros. The green line represents the Fisher zeros in the thermodynamic limit. The discrete Fisher zeros for $L=16$ roughly lie on the continuous curve in the thermodynamic limit. (b) The red (blue) dots represent the results of the real (imaginary) parts of the partition function obtained using HOTRG for $L=16$, while the red (blue) lines represent the results obtained using ED. The Fisher zeros obtained by the two methods are in complete agreement.
	\label{fig:circle_comp}}
\end{figure}

The original Suzuki solution~\cite{Suzuki1976S} of the partition function of the one-dimensional transverse field Ising Model (1DTFIM) is 
\begin{equation}
	\begin{split}
	Z_{\rm{su}}& = \frac{1}{2}
	\left[
	\prod_{k=1}^L 2\cosh(\beta\epsilon_{2k})+\prod_{k=1}^L 2\sinh(\beta\epsilon_{2k})\right.\\
	&\left.+\prod_{k=1}^L 2\cosh(\beta\epsilon_{2k-1})+\prod_{k=1}^L 2\sinh(\beta\epsilon_{2k-1})
	\right],
	\end{split}
	\label{eq:Zsu1}
\end{equation}
where $\epsilon_k = [1 + g^2 - 2g\cos(\pi k/L)]^{1/2}$ and the Kramers-Wannier (KW)~\cite{KW1941S} is fully manifested. Specifically, Eq.~\ref{eq:Zsu1} remains invariant under the transformation $g\rightarrow 1/g$, $\beta\rightarrow\beta g$. Consequently, the Fisher zero results for $g > 1$ can be easily derived by mapping to the dual space ($g < 1$). Therefore, zero lines will reappear for $g > 1$ to characterize the dual spin flip excitations, similar to those for domain walls. Figure~\ref{fig:circle_comp}(a) shows the results of Fisher zeros at $g=1.5$ for $L=16$ and in the thermodynamic limit, according to the Suzuki solution and the exact formula given in Ref.~\cite{liu2023CPLS}, 
thereby illustrating the KW duality behavior between spin flips and domain walls on the two sides of the QCP in the complex $\beta$ plane.

To verify whether the physical picture at $g>1$ obtained from the Suzuki solution is correct, we employ two precise numerical methods, exact diagonalization (ED) and tensor network calculations, to obtain Fisher zeros solutions for the same small system with $L=16$. In the tensor network calculations, we first construct a 2D lattice of size $L\times n$ from the 1D system by using Trotter decomposition with $\beta=\tau n$, then solve the partition function using the higher-order tensor renormalization group (HOTRG) method~\cite{XieHOTRGS}, 
an efficient tensor network algorithm capable of handling sign problems in the complex $\beta$ plane~\cite{Zou2014PRDS}. 
In our HOTRG calculations, $n=$1000, and the bond dimension $D_b$ describing the entanglement properties is set to 32. We find that the ED and the HOTRG calculations provide consistent results~[Fig.~\ref{fig:circle_comp}(b)], but they are entirely inconsistent with the picture from the Suzuki solution with the same system size $L=16$. In the latter two numerical results, there is no reappearance of Fisher zero lines in the complex $\beta$ plane; instead, only localized Fisher zero structures are found~[Fig.~\ref{fig:circle_comp}(b)], indicating the presence of closed zero curves in the thermodynamic limit. This implies that a new description is needed for spin flip excitations in the quantum disordered phase on the complex $\beta$ plane, as the Fisher zero lines strictly adhering to KW duality according to the Suzuki solution will no longer apply.

\subsection{Modified Suzuki solution on quantum disordered side}
We then discuss the reasons for the discrepancy between results from the  Suzuki solution and those from the unbiased numerical calculations. Prior to deriving Eq.~\ref{eq:Zsu1}, anisotropic Onsager/Kaufman results are used, with the Trotter step $n$ tends to infinity:
\begin{equation}
	\begin{split}
	Z = &\lim_{n\rightarrow\infty} \frac{1}{2}
	\left[
	\prod_{k=1}^L 2\cosh\left(\frac{n}{2}\gamma_{2k}\right)+\prod_{k=1}^L 2\sinh\left(\frac{n}{2}\gamma_{2k}\right)\right.\\
	&\left.+\prod_{k=1}^L 2\cosh\left(\frac{n}{2}\gamma_{2k-1}\right)+\prod_{k=1}^L 2\sinh\left(\frac{n}{2}\gamma_{2k-1}\right)
	\right],
	\end{split}
	\label{eq:Zlim}
\end{equation}
where $\gamma_k$ is given by 
\begin{equation}
  \begin{split}
  \cosh\gamma_k &= \cosh\left(\frac{2}{n}\beta g\right)\cosh\left(\frac{2}{n}\beta\right)\\
  &-\sinh\left(\frac{2}{n}\beta g\right)\sinh\left(\frac{2}{n}\beta\right)\cos\left(\frac{\pi k}{L}\right).
  \end{split}
\end{equation}
The two largest eigenvalues $\lambda_{\pm}$ provide the majority contribution to the partition function, with $Z\sim\lambda_+^{n}+\lambda_-^{n}$ and 
\begin{equation}
\begin{split}
  \lambda_+=&\exp\left[\frac{1}{2}\left(\gamma_1+\gamma_3+\cdots+\gamma_{2L-1}\right)\right],\\
   \lambda_-=&\exp\left[\frac{1}{2}\left(\gamma_2+\gamma_4+\cdots+\gamma_{2L}\right)\right].
\end{split}
\end{equation}

The reason why $Z$ of the 1DTFIM should be different on the two sides of the QCP is similar to why the partition function of the 2D classical Ising model differs on the two side of the critical temperature~\cite{Kaufman1949S}. On the ordered side ($g<1$), the ground state of the 1DTFIM has a degeneracy of two: $|\uparrow\uparrow\cdots\uparrow\rangle$ and $|\downarrow\downarrow\cdots\downarrow\rangle$. To recover this property at $\beta\rightarrow\infty$, it must satisfy the condition $\lambda_+=\lambda_-$ in the thermodynamic limit. On the other hand, on the quantum disordered side ($g<1$), the ground state has a degeneracy of only one $|\rightarrow\rightarrow\cdots\rightarrow\rangle$. Correspondingly, this requires $\lambda_+<\lambda_-$ in $Z$. At $n\rightarrow\infty$ and $L\rightarrow\infty$, $\gamma_k=2\beta\epsilon_k/n$, with $\gamma_{2k-1}$ and $\gamma_{2k}$ are approximately equal for $k<L$. At $k=L$, however, $\gamma_{2k}$ has two solutions. In fact, the relationship between $\lambda_+$ and $\lambda_-$ due to degeneracy on  two sides of the QCP can be adjusted by choosing the sign of $\gamma_{2L}$. By selecting $\gamma_{2L}=2\beta(g_c-g)/n$, the relations $\lambda_-/\lambda_+=1$ ($g<g_c$) and $\lambda_-/\lambda_+<1$ ($g>g_c$) are achieved. In the original Suzuki solution, $\gamma_{2L}=2\beta|g_c-g|/n$, leading to the reappearance of zero lines at $g>g_c$ and creating a fake illusion of KW duality on the complex $\beta$ plane. Therefore, the exact Suzuki solution at $g>g_c$ need to modify as  
\begin{equation}
	\begin{split}
	Z_{\rm{su}}'& = \frac{1}{2}
	\left[
	\prod_{k=1}^L 2\cosh(\beta\epsilon_{2k})-\prod_{k=1}^L 2\sinh(\beta\epsilon_{2k})\right.\\
	&\left.+\prod_{k=1}^L 2\cosh(\beta\epsilon_{2k-1})+\prod_{k=1}^L 2\sinh(\beta\epsilon_{2k-1})
	\right].
	\end{split}
	\label{eq:Zsu2}
\end{equation}

\begin{figure}[t]
    \centering
    \includegraphics[width=0.48\textwidth]{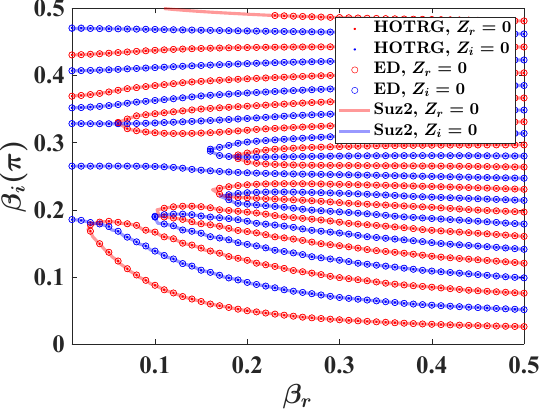}
    \caption{Fisher zeros obtained from three different methods for $L=16$ and $g=1.5$. The results obtained using the modified Suzuki solution (lines) perfectly match those from HOTRG (dots) and ED (circles).}
    \label{fig:compare3}
\end{figure}

In Fig.~\ref{fig:compare3}, the modified Suzuki solution Eq.~\ref{eq:Zsu2} yields the correct picture of Fisher zeros, consistent with whose from other two numerical results.

\section{Structure of Fisher zeros}

\subsection{Finding Fisher zeros from the analytical solution of $Z$ for the 1DTFIM}

\begin{figure}[h]
    \includegraphics[width=1.0\textwidth]{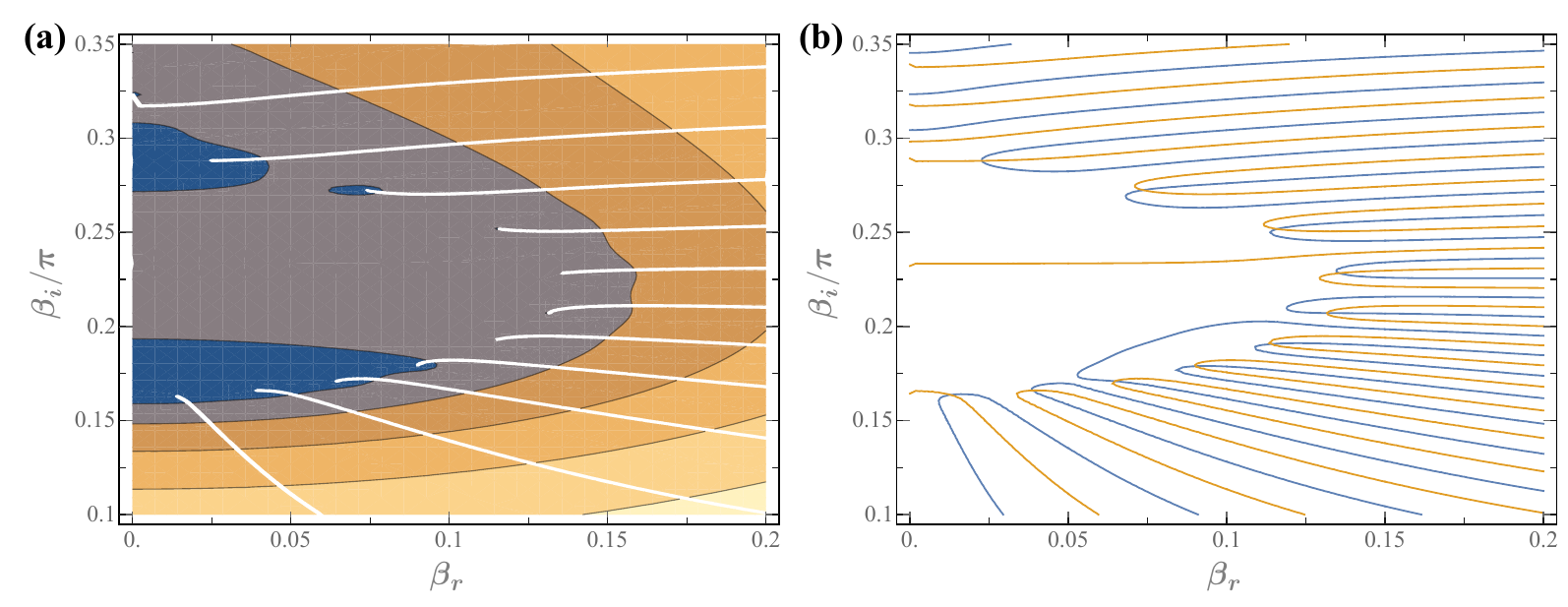}
    \caption{Locating Fisher zeros. (a) The contour lines of ${\rm Re}f$. The white
lines indicate the branch cuts, which terminate the Fisher zeros at the starting points ($L=32$,$g=1.8$). (b) Zeros of the same system determined by the intersections of ${\rm Re}(Z)=0$ and ${\rm Im}(Z)=0$ from the exact solution.
    \label{fig:freeE}}
\end{figure}

With the form of $Z$ determined, we now sketch how Fisher zeros are located and visualized.
To analytically continue $Z$, we simply allow $\beta$ to be complex, $\beta=\beta_r+i\beta_i$, in Eq.~(2) in the main text.
On the complex $\beta$ plane, $Z(\beta)$ is not a polynomial but an entire function of $\beta$.
Following Fisher~\cite{fisher1965statisticalS}, we can factorize $Z$ into
$
Z(\beta) = e^{h(\beta)}\prod_j(\beta -z_j)
$
where $h$ is an entire function, and the multiplicity of $z_j$, the $j$-th root of $Z=0$, is assumed to be 1.
Define the free energy (up to a factor of $\beta$) as 
$f=-\ln Z = - h - f_{na}$, where the non-analytical part 
$
f_{na}=\sum_j \ln (\beta -z_j)
$
develops singularities at $\beta =z_j$ and is completely determined by 
the Fisher zeros $\{z_j\}$. Let $f_{na}=\phi + i\psi$, then real part $\phi=\sum_i \ln|\beta-z_j|$
can be interpreted as the electric potential due to a set of point charges at $\{z_j\}$~\cite{bena2005statisticalS}.
The electric field lines stem out of these Fisher charges and correspond to the contour line of the imaginary part $\psi$~\cite{bena2005statisticalS}. 
A branch cut of $f$ is developed at each Fisher zero. Thus the contour plot of the real and imaginary part of 
$f$ can serve as an intuitive way to locate the Fisher zeros, matching the exact solution (Fig.~\ref{fig:freeE}). 
Note that in Fig.~\ref{fig:freeE}, due to the mirror symmetry of Fisher zeros
with respect to the $\beta_i$ axis,
only the Fisher zeros at the first quadrant is displayed. 
In the thermodynamic limit, if the Fisher zeros
come close to each other, the charge 
density $\rho(\beta)=\sum_j \delta(\beta-z_j)$ may become a continuous function.

\subsection{Velocity of zeros of the 1DTFIM in the thermodynamic limit}

In the thermodynamic limit, the solution of $Z_{\rm su}=0$ at $g<1$ can be written as~\cite{liu2023CPLS}
\begin{equation}
\tilde{S}=\int_0^\pi\log |\tanh(\beta \epsilon_q)|^2dq=0,
\label{eq:exthem}
\end{equation}
where $\epsilon_q=(1+g^2-2g\cos q)^{1/2}$. Due to the different signs of the $\sinh$ terms containing $\epsilon_{2k}$ and $\epsilon_{2k-1}$, $\tilde{S}$ at $g>1$ in the thermodynamic limit is difficult to express in a simple form. However, we can still obtain an asymptotic solution in the thermodynamic limit using the solution for large 
$L$. Therefore, we formally represent the solution for the zeros as $\tilde{S}(\beta,g)=0$. When $g$ undergoes a slight change to $g'$, the zero moves to $\beta'$ in the complex plane, resulting in $\tilde{S}(\beta',g')=0$. By performing a Taylor expansion of $\tilde{S}$ at $(\beta,g)$ up to the first order, we obtain the velocity of the zeros' movement in the complex $\beta$ plane with respect to changes in $g$:
\begin{equation}
v_g=\frac{\partial\beta}{\partial g}=-\left(\frac{\partial\tilde{S}}{\partial g}\right)_\beta/\left(\frac{\partial\tilde{S}}{\partial\beta}\right)_g,
\end{equation}
which is similar to the Gr\"uneisen parameter. At the critical point, the zeros on the closed Fisher curve have a finite speed, while on the open Fisher curve, the speed of the zeros diverges at large $\beta$ and in the thermodynamic limit. These conclusions are consistent with the behavior of the Gr\"uneisen parameter at the critical point~\cite{Zhang2019prlS,Zhang2023cplS}.

\subsection{Fisher zeros in other spin models}
\begin{figure}[t]
    \includegraphics[width=1.0\textwidth]{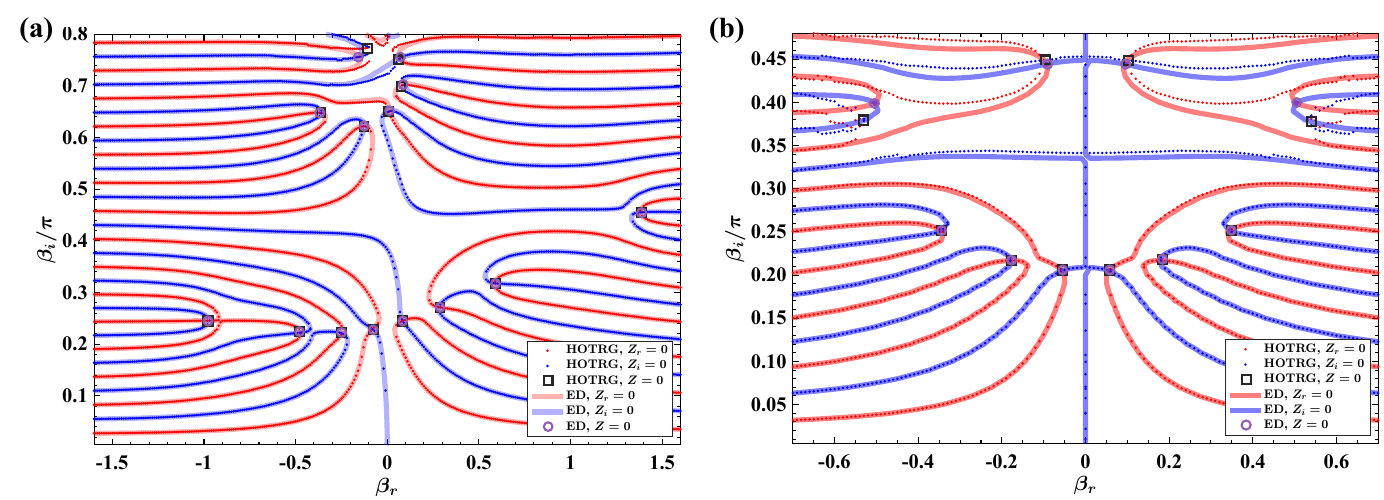}
    \caption{The Fisher zeros of two different models, obtained from ED and tensor network calculation at $L=16$, are consistent. (a) ANNNI model with $\kappa =0.1$, $g=0.5$ and $D_b=30$ (b) XY model with $\gamma=0.01$ and $D_b=50$.
    \label{fig:morezeros}}
\end{figure}

We first consider the ANNNI model
\begin{equation}
H=-\sum_{i=1}^{L} \left(\sigma_i^z\sigma_{i+1}^z -\kappa\sigma_i^z\sigma_{i+2}^z+g\sigma_i^x\right),
\end{equation}
where $\kappa$ is the next-nearest-neighbor antiferromagnetic coupling and $g$ is the transverse field. In the case where  $\kappa$ and $g$ are relatively small, the excitations can still be described by domain walls. Results of the Fisher zeros using ED and tensor network methods consistently show that the open line-shaped zeros [Fig.~\ref{fig:morezeros}(a)], which describe domain wall excitations in the 1DTFIM, still exist. The introduction of $\kappa$ only causes this line to tilt.

We further consider the XY model
\begin{equation}
H=\sum_{i=1}^{L}[(1+\gamma)\sigma_i^x\sigma_{i+1}^x+(1-\gamma)\sigma_i^y\sigma_{i+1}^y],
\end{equation}
where at $\gamma=0$, it transforms into the gapless XX model. We choose a small $g$ at $g=0.01$ to perform Fisher zero calculations using ED and tensor network methods. The tensor network results are accurate for most zero points, except for some deviations from the ED results in the relatively larger region of $\beta_i$. Since the system is approaching the gapless point, more precise tensor network calculations require a larger $D_b$. We find that in the complex $\beta$ plane near $\beta_r$ axis, the open line-shaped zeros completely disappear, leaving only zeros that form closed loops. This is similar to the behavior near the QCP in the 1DTFIM.

The results of the Fisher zeros in these two models indicate that the zero structures exhibited by the 1DTFIM and their relation to quantum criticality have a certain generality. By analyzing the variation of zeros with parameters in these and other models, we can explore the properties of various many-body systems.

\subsection{Scaling of the Zeros near QCP}

\begin{figure}[h]
    \includegraphics[width=1.0\textwidth]{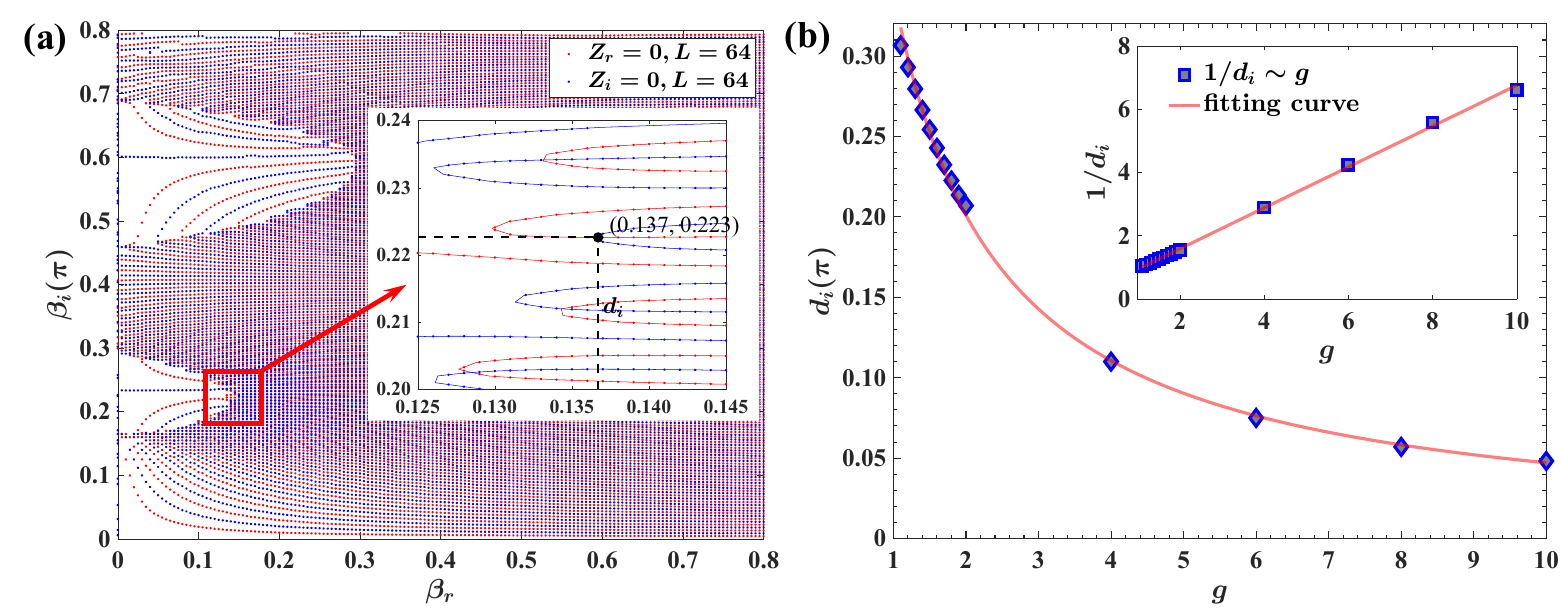}
    \caption{The Fisher zeros obtained from HOTRG at $L=64$. (a) An example of the Results at $g=1.8$. The intersections of red (blue) points determine the location of the Fisher zeros. The zeros are densely distributed and approximate form closed curve structures. The inset illustrates the distance $d_i$ from the rightmost point of the first approximate curve to the real $\beta$ axis. (b) The distance $d_i$ obtained at different $g$ values exhibits an inverse proportionality with $g$. The inset depicts the fit with $0.65(g+0.45)$. 
    \label{fig:FisherZerosCircles}}
\end{figure}

As mentioned in the text, the zeros of the partition function can be located numerically by tensor network calculation. We can determine the positions of these zeros under fixed $\beta_r$ by scanning $Z$ with respect to $\beta_i$. Figure~\ref{fig:FisherZerosCircles}(a) illustrates several examples at $L=64$ and $g=1.8$. Our numerical results from HOTRG are consistent with the exact solutions. 
As $L$ increases, the zeros indeed become denser and tend toward some closed curves. By comparing the approximate zero curves for different $g$ values, we find that as $g$ increases, the size of the closed curve gradually decreases and approaches the real axis. Since the inverse of imaginary part of $\beta$ ($1/\beta_i$) provides the energy scale of quantum fluctuations~\cite{liu2023CPLS}, 
we choose the distance $d_i$ from the rightmost point of the zero curve closest to the real axis to the real axis for scaling. We find that as $g$ increases, this distance $d_i$ decreases inversely with $g$ [Fig.~\ref{fig:FisherZerosCircles}(b)]. This scaling relationship is consistent with the linear relation between the spin flip excitation gap $\Delta$ and $g$. Therefore, on the quantum disordered side, the size of closed Fisher zero curves can be used to characterize spin flip excitations, while Fisher zero lines used to characterize domain walls on the ordered side.
The different manifestations of excitations on Fisher zeros on the two sides of the QCP indicate that the KW duality does not necessarily enforce similarities in the behaviors of the renormalized classical side and the quantum disorder side in the complex $\beta$ plane.

\section{Comparing the Suzuki solution and CFT at the quantum critical point}

In the thermodynamic limit $L\rightarrow\infty$ and at the zero temperature $\beta\rightarrow\infty$, the partition function formula for the Ising critical universality described by Conformal Field Theory (CFT) can be derived from the Suzuki formula.
At the critical point $g=g_c=1$, the relationship $\epsilon_k=2\sin(\pi k/2L)$ holds. Therefore, $\epsilon_{2L}$ in the second term of the Suzuki solution vanishes, and $Z$ can be rewrote as
\begin{eqnarray*}
Z=&&\frac{1}{2}\left[
e^{2\beta\sum_{k=1}^L\sin(\pi k/L)}\prod_{k=1}^L\left(1+e^{-4\beta\sin(\pi k/L)}\right)
+e^{2\beta\sum_{k=1}^L\sin[\pi(k-1/2)/L]}\prod_{k=1}^L\left(1+e^{-4\beta\sin[\pi (k-1/2)/L]}\right) \right.
\\
&&\left.+e^{2\beta\sum_{k=1}^L\sin[\pi(k-1/2)/L]}\prod_{k=1}^L\left(1-e^{-4\beta\sin[\pi (k-1/2)/L]}\right)\right].
\label{eq:form1}
\end{eqnarray*}

At large $L$, using
\[
\sum_{k=1}^L\sin\frac{\pi k}{L}=\cot\frac{\pi}{2L}=\frac{2L}{\pi}-\frac{\pi}{6L}+O(\frac{1}{L^3}),
\] 
and considering that at $k\sim 0$, $\sin(\pi k/L)=\sin[\pi(L-k)/L]\sim\pi k/L$ and at $k\sim L/2$, $e^{-4\beta\sin(\pi k/L)}\sim e^{-4\beta}$ is tiny at large $\beta$, we have
\[
\prod_{k=1}^L\left(1+e^{-4\beta\sin(\pi k/L)}\right)=2\prod_{k=1}^\infty\left( 1+e^{-4(\beta/L)\pi k)}\right).
\]
Therefore, the first term in the Suzuki solution is then
\[
e^{4L^2(\beta/L)/\pi}2\tilde{q}^{\frac{1}{12}}\prod_{k=1}^\infty(1+\tilde{q}^k)^2,
\]
where $q\equiv e^{-2\pi\beta/L}$ and $\tilde{q}=q^2$.

By using
\[
\sum_{k=1}^L\sin\frac{\pi}{L}(k-1/2)=\csc\frac{\pi}{2L}=\frac{2L}{\pi}+\frac{\pi}{12L}+O(\frac{1}{L^3}),
\] 
and treating the second and third terms in the Suzuki solution similarly, the full partition function can be expressed as
\[
Z=e^{4L^2(\beta/L)/\pi}\left[2\tilde{q}^{\frac{1}{12}}\prod_{k=1}^\infty(1+\tilde{q}^k)^2+\tilde{q}^{-\frac{1}{24}}\prod_{k=1}^\infty(1+\tilde{q}^{k-1/2})^2+
\tilde{q}^{-\frac{1}{24}}\prod_{k=1}^\infty(1-\tilde{q}^{k-1/2})^2\right].
\]

\begin{figure}[t]
    \centering
    \includegraphics[width=1\textwidth]{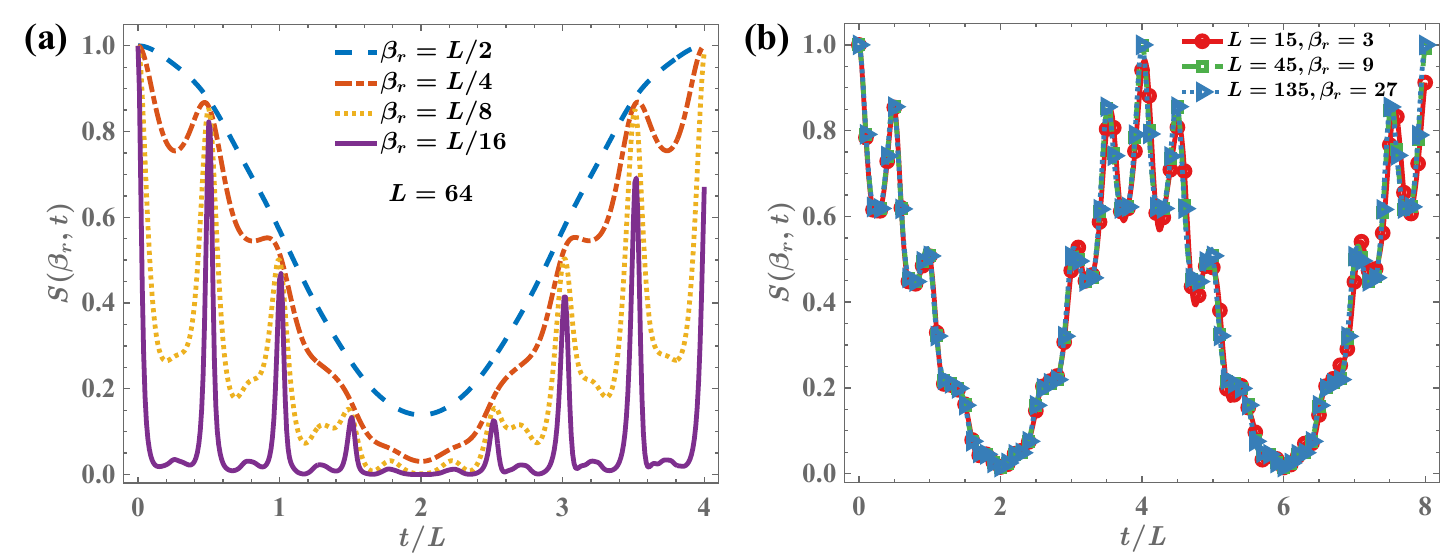}
    \caption{The oscillatory behavior of $S(\beta_r,t)$ at the QCP with different $L$ and $\beta_r$. (a) Fine structures of $S$ show the same periodicity $t/L=4$. (b) Example of self similarity of $S$ with $n=3$.}
    \label{fig:simlar}
\end{figure}

In CFT, $\theta(\tau)$ function and Dedekind $\eta(\tau)$ function are usually used, with
\begin{eqnarray*}
\theta_2(\tau)=&&2q^{\frac{1}{8}}\prod_{k=1}^\infty(1-q^k)(1+q^k)^2,\\
\theta_3(\tau)=&&\prod_{k=1}^\infty(1-q^k)(1+q^{k-1/2})^2,\\
\theta_4(\tau)=&&\prod_{k=1}^\infty(1-q^k)(1+q^{k-1/2})^2,\\
\eta(\tau)=&&q^{\frac{1}{24}}\prod_{k=1}^\infty(1-q^k),
\end{eqnarray*}
where $\tau$ is the modular parameter and $q=e^{2\pi i\tau}$. For 1DTFIM, by defining $\tau\equiv i\beta/L$, the partition function is then
\[
Z=\frac{1}{2}e^{-i4\tau L^2/\pi}\left[\frac{\theta_2(2\tau)}{\eta(2\tau)}+
\frac{\theta_3(2\tau)}{\eta(2\tau)}+\frac{\theta_4(2\tau)}{\eta(2\tau)}\right],
\]
which is consistent with the CFT result of the 2D classical Ising model~\cite{CFTbookS}, except for the substitution of $\tau$ with $2\tau$. This $2\tau$ term in $Z$ causes the quantum recurrence time $t^*$ to be halved compared to the CFT result, so that $t^*=4L$ for the 1DTFIM [Fig.~\ref{fig:simlar}(a)]. Actually, this does not contradict the result $t^*=8L$ in Ref.~\cite{NiveditaPRE2020S}. In Ref.~\cite{NiveditaPRE2020S}, the Hamiltonian for the 1DTFIM includes a factor of $1/2$, which, after the Jordan-Wigner transformation, results in a form with a single fermionic excitation, thereby matching the result with the Ising CFT. In fermionic representation, the 1DTFIM form in this paper is then corresponds to two fermionic excitations. Since the period depends on the system's energy scale, doubling the energy exactly halves the period.

\section{Self-similarity of $S$}
Self-similarity is a fundamental characteristic of quantum critical points, reflecting the scale-invariant nature of quantum fluctuations at these points. The relationship $S(\beta_r,t)_L\approx S(n\beta_r,nt)_{nL}$ at $g_c=1$ provides a classic example of how systems at different scales can exhibit similar behavior. This property can be used for RG analysis, a powerful tool in studying critical phenomena. Figure~\ref{fig:simlar}(b) illustrates the self-similar behavior of $S$ using $n=3$ as an example. The oscillatory behavior of $S$ at different $L$ values not only exhibits the same periodicity but also shows nearly perfect overlap, demonstrating the self-similar nature. The self-similarity implies that the critical exponents, which characterize the behavior of physical quantities near the QCP, are scale-invariant.  

Similar to correlation functions, $S$ can be used for two-lattice matching RG analysis. In-depth analysis of the RG flow derived from this self-similarity relationship sheds light on the underlying mechanisms governing quantum phase transitions. By matching $S(\beta)_L$=$S(\beta')_{L'}$ from the system with size $L$ to $L'$, we can define an RG flow $\beta\rightarrow\beta'$ in the complex plane. This flow starts from large $\beta$ region and can reach small $\beta$, until the singularity near the Fisher zeros disrupt the self-similar behavior. By observing the trajectory of $\beta$ in the complex plane as it transforms into $\beta'$, one can map the evolution of the system across different scales. This mapping not only validates the theoretical predictions of critical phenomena but also provides a quantitative tool to explore the dynamics of quantum critical systems.


\begin{thebibliography}{49}%
\makeatletter
\providecommand \@ifxundefined [1]{%
 \@ifx{#1\undefined}
}%
\providecommand \@ifnum [1]{%
 \ifnum #1\expandafter \@firstoftwo
 \else \expandafter \@secondoftwo
 \fi
}%
\providecommand \@ifx [1]{%
 \ifx #1\expandafter \@firstoftwo
 \else \expandafter \@secondoftwo
 \fi
}%
\providecommand \natexlab [1]{#1}%
\providecommand \enquote  [1]{``#1''}%
\providecommand \bibnamefont  [1]{#1}%
\providecommand \bibfnamefont [1]{#1}%
\providecommand \citenamefont [1]{#1}%
\providecommand \href@noop [0]{\@secondoftwo}%
\providecommand \href [0]{\begingroup \@sanitize@url \@href}%
\providecommand \@href[1]{\@@startlink{#1}\@@href}%
\providecommand \@@href[1]{\endgroup#1\@@endlink}%
\providecommand \@sanitize@url [0]{\catcode `\\12\catcode `\$12\catcode
  `\&12\catcode `\#12\catcode `\^12\catcode `\_12\catcode `\%12\relax}%
\providecommand \@@startlink[1]{}%
\providecommand \@@endlink[0]{}%
\providecommand \url  [0]{\begingroup\@sanitize@url \@url }%
\providecommand \@url [1]{\endgroup\@href {#1}{\urlprefix }}%
\providecommand \urlprefix  [0]{URL }%
\providecommand \Eprint [0]{\href }%
\providecommand \doibase [0]{https://doi.org/}%
\providecommand \selectlanguage [0]{\@gobble}%
\providecommand \bibinfo  [0]{\@secondoftwo}%
\providecommand \bibfield  [0]{\@secondoftwo}%
\providecommand \translation [1]{[#1]}%
\providecommand \BibitemOpen [0]{}%
\providecommand \bibitemStop [0]{}%
\providecommand \bibitemNoStop [0]{.\EOS\space}%
\providecommand \EOS [0]{\spacefactor3000\relax}%
\providecommand \BibitemShut  [1]{\csname bibitem#1\endcsname}%
\let\auto@bib@innerbib\@empty
\bibitem [{\citenamefont {Yang}\ and\ \citenamefont {Lee}(1952)}]{LeeYang1}%
  \BibitemOpen
  \bibfield  {author} {\bibinfo {author} {\bibfnamefont {C.~N.}\ \bibnamefont
  {Yang}}\ and\ \bibinfo {author} {\bibfnamefont {T.~D.}\ \bibnamefont {Lee}},\
  }\bibfield  {title} {\bibinfo {title} {Statistical theory of equations of
  state and phase transitions. i. theory of condensation},\ }\href
  {https://doi.org/10.1103/PhysRev.87.404} {\bibfield  {journal} {\bibinfo
  {journal} {Phys. Rev.}\ }\textbf {\bibinfo {volume} {87}},\ \bibinfo {pages}
  {404} (\bibinfo {year} {1952})}\BibitemShut {NoStop}%
\bibitem [{\citenamefont {Lee}\ and\ \citenamefont {Yang}(1952)}]{LeeYang2}%
  \BibitemOpen
  \bibfield  {author} {\bibinfo {author} {\bibfnamefont {T.~D.}\ \bibnamefont
  {Lee}}\ and\ \bibinfo {author} {\bibfnamefont {C.~N.}\ \bibnamefont {Yang}},\
  }\bibfield  {title} {\bibinfo {title} {Statistical theory of equations of
  state and phase transitions. ii. lattice gas and ising model},\ }\href
  {https://doi.org/10.1103/PhysRev.87.410} {\bibfield  {journal} {\bibinfo
  {journal} {Phys. Rev.}\ }\textbf {\bibinfo {volume} {87}},\ \bibinfo {pages}
  {410} (\bibinfo {year} {1952})}\BibitemShut {NoStop}%
\bibitem [{\citenamefont {Fisher}\ and\ \citenamefont
  {Brittin}(1965)}]{fisher1965statistical}%
  \BibitemOpen
  \bibfield  {author} {\bibinfo {author} {\bibfnamefont {M.}~\bibnamefont
  {Fisher}}\ and\ \bibinfo {author} {\bibfnamefont {W.}~\bibnamefont
  {Brittin}},\ }\bibfield  {title} {\bibinfo {title} {Statistical physics, weak
  interactions, field theory},\ }\href@noop {} {\bibfield  {journal} {\bibinfo
  {journal} {Lectures in Theoretical Physics (Boulder: University of Colorado
  Press) vol VIIC}\ } (\bibinfo {year} {1965})}\BibitemShut {NoStop}%
\bibitem [{\citenamefont {Bena}\ \emph {et~al.}(2005)\citenamefont {Bena},
  \citenamefont {Droz},\ and\ \citenamefont {Lipowski}}]{bena2005statistical}%
  \BibitemOpen
  \bibfield  {author} {\bibinfo {author} {\bibfnamefont {I.}~\bibnamefont
  {Bena}}, \bibinfo {author} {\bibfnamefont {M.}~\bibnamefont {Droz}},\ and\
  \bibinfo {author} {\bibfnamefont {A.}~\bibnamefont {Lipowski}},\ }\bibfield
  {title} {\bibinfo {title} {Statistical mechanics of equilibrium and
  nonequilibrium phase transitions: the yang--lee formalism},\ }\href@noop {}
  {\bibfield  {journal} {\bibinfo  {journal} {International Journal of Modern
  Physics B}\ }\textbf {\bibinfo {volume} {19}},\ \bibinfo {pages} {4269}
  (\bibinfo {year} {2005})}\BibitemShut {NoStop}%
\bibitem [{\citenamefont {Kist}\ \emph {et~al.}(2021)\citenamefont {Kist},
  \citenamefont {Lado},\ and\ \citenamefont {Flindt}}]{Flindt2021prr}%
  \BibitemOpen
  \bibfield  {author} {\bibinfo {author} {\bibfnamefont {T.}~\bibnamefont
  {Kist}}, \bibinfo {author} {\bibfnamefont {J.~L.}\ \bibnamefont {Lado}},\
  and\ \bibinfo {author} {\bibfnamefont {C.}~\bibnamefont {Flindt}},\
  }\bibfield  {title} {\bibinfo {title} {Lee-yang theory of criticality in
  interacting quantum many-body systems},\ }\href
  {https://doi.org/10.1103/PhysRevResearch.3.033206} {\bibfield  {journal}
  {\bibinfo  {journal} {Phys. Rev. Res.}\ }\textbf {\bibinfo {volume} {3}},\
  \bibinfo {pages} {033206} (\bibinfo {year} {2021})}\BibitemShut {NoStop}%
\bibitem [{\citenamefont {Liu}\ \emph {et~al.}(2023)\citenamefont {Liu},
  \citenamefont {Lv}, \citenamefont {Yang},\ and\ \citenamefont
  {Zou}}]{liu2023CPL}%
  \BibitemOpen
  \bibfield  {author} {\bibinfo {author} {\bibfnamefont {Y.}~\bibnamefont
  {Liu}}, \bibinfo {author} {\bibfnamefont {S.}~\bibnamefont {Lv}}, \bibinfo
  {author} {\bibfnamefont {Y.}~\bibnamefont {Yang}},\ and\ \bibinfo {author}
  {\bibfnamefont {H.}~\bibnamefont {Zou}},\ }\bibfield  {title} {\bibinfo
  {title} {Signatures of quantum criticality in the complex inverse temperature
  plane},\ }\href@noop {} {\bibfield  {journal} {\bibinfo  {journal} {Chinese
  Physics Letters}\ }\textbf {\bibinfo {volume} {40}},\ \bibinfo {pages}
  {050502} (\bibinfo {year} {2023})}\BibitemShut {NoStop}%
\bibitem [{\citenamefont {Sachdev}(2011)}]{Sachdevbook}%
  \BibitemOpen
  \bibfield  {author} {\bibinfo {author} {\bibfnamefont {S.}~\bibnamefont
  {Sachdev}},\ }\href@noop {} {\emph {\bibinfo {title} {Quantum Phase
  Transitions}}}\ (\bibinfo  {publisher} {Cambridge University
  Press,Cambridge},\ \bibinfo {year} {2011})\BibitemShut {NoStop}%
\bibitem [{\citenamefont {Kramers}\ and\ \citenamefont
  {Wannier}(1941)}]{KW1941}%
  \BibitemOpen
  \bibfield  {author} {\bibinfo {author} {\bibfnamefont {H.~A.}\ \bibnamefont
  {Kramers}}\ and\ \bibinfo {author} {\bibfnamefont {G.~H.}\ \bibnamefont
  {Wannier}},\ }\bibfield  {title} {\bibinfo {title} {Statistics of the
  two-dimensional ferromagnet. part i},\ }\href
  {https://doi.org/10.1103/PhysRev.60.252} {\bibfield  {journal} {\bibinfo
  {journal} {Phys. Rev.}\ }\textbf {\bibinfo {volume} {60}},\ \bibinfo {pages}
  {252} (\bibinfo {year} {1941})}\BibitemShut {NoStop}%
\bibitem [{\citenamefont {Fisher}\ \emph {et~al.}(2023)\citenamefont {Fisher},
  \citenamefont {Khemani}, \citenamefont {Nahum},\ and\ \citenamefont
  {Vijay}}]{fisher2023random}%
  \BibitemOpen
  \bibfield  {author} {\bibinfo {author} {\bibfnamefont {M.~P.}\ \bibnamefont
  {Fisher}}, \bibinfo {author} {\bibfnamefont {V.}~\bibnamefont {Khemani}},
  \bibinfo {author} {\bibfnamefont {A.}~\bibnamefont {Nahum}},\ and\ \bibinfo
  {author} {\bibfnamefont {S.}~\bibnamefont {Vijay}},\ }\bibfield  {title}
  {\bibinfo {title} {Random quantum circuits},\ }\href@noop {} {\bibfield
  {journal} {\bibinfo  {journal} {Annual Review of Condensed Matter Physics}\
  }\textbf {\bibinfo {volume} {14}},\ \bibinfo {pages} {335} (\bibinfo {year}
  {2023})}\BibitemShut {NoStop}%
\bibitem [{\citenamefont {Heyl}\ \emph {et~al.}(2013)\citenamefont {Heyl},
  \citenamefont {Polkovnikov},\ and\ \citenamefont
  {Kehrein}}]{DynamicalPT2013PRL}%
  \BibitemOpen
  \bibfield  {author} {\bibinfo {author} {\bibfnamefont {M.}~\bibnamefont
  {Heyl}}, \bibinfo {author} {\bibfnamefont {A.}~\bibnamefont {Polkovnikov}},\
  and\ \bibinfo {author} {\bibfnamefont {S.}~\bibnamefont {Kehrein}},\
  }\bibfield  {title} {\bibinfo {title} {Dynamical quantum phase transitions in
  the transverse-field ising model},\ }\href
  {https://doi.org/10.1103/PhysRevLett.110.135704} {\bibfield  {journal}
  {\bibinfo  {journal} {Phys. Rev. Lett.}\ }\textbf {\bibinfo {volume} {110}},\
  \bibinfo {pages} {135704} (\bibinfo {year} {2013})}\BibitemShut {NoStop}%
\bibitem [{\citenamefont {Li}\ \emph {et~al.}(2018)\citenamefont {Li},
  \citenamefont {Chen},\ and\ \citenamefont {Fisher}}]{PhysRevB.98.205136}%
  \BibitemOpen
  \bibfield  {author} {\bibinfo {author} {\bibfnamefont {Y.}~\bibnamefont
  {Li}}, \bibinfo {author} {\bibfnamefont {X.}~\bibnamefont {Chen}},\ and\
  \bibinfo {author} {\bibfnamefont {M.~P.~A.}\ \bibnamefont {Fisher}},\
  }\bibfield  {title} {\bibinfo {title} {Quantum zeno effect and the many-body
  entanglement transition},\ }\href
  {https://doi.org/10.1103/PhysRevB.98.205136} {\bibfield  {journal} {\bibinfo
  {journal} {Phys. Rev. B}\ }\textbf {\bibinfo {volume} {98}},\ \bibinfo
  {pages} {205136} (\bibinfo {year} {2018})}\BibitemShut {NoStop}%
\bibitem [{\citenamefont {Skinner}\ \emph {et~al.}(2019)\citenamefont
  {Skinner}, \citenamefont {Ruhman},\ and\ \citenamefont
  {Nahum}}]{PhysRevX.9.031009}%
  \BibitemOpen
  \bibfield  {author} {\bibinfo {author} {\bibfnamefont {B.}~\bibnamefont
  {Skinner}}, \bibinfo {author} {\bibfnamefont {J.}~\bibnamefont {Ruhman}},\
  and\ \bibinfo {author} {\bibfnamefont {A.}~\bibnamefont {Nahum}},\ }\bibfield
   {title} {\bibinfo {title} {Measurement-induced phase transitions in the
  dynamics of entanglement},\ }\href
  {https://doi.org/10.1103/PhysRevX.9.031009} {\bibfield  {journal} {\bibinfo
  {journal} {Phys. Rev. X}\ }\textbf {\bibinfo {volume} {9}},\ \bibinfo {pages}
  {031009} (\bibinfo {year} {2019})}\BibitemShut {NoStop}%
\bibitem [{\citenamefont {Chan}\ \emph {et~al.}(2019)\citenamefont {Chan},
  \citenamefont {Nandkishore}, \citenamefont {Pretko},\ and\ \citenamefont
  {Smith}}]{PhysRevB.99.224307}%
  \BibitemOpen
  \bibfield  {author} {\bibinfo {author} {\bibfnamefont {A.}~\bibnamefont
  {Chan}}, \bibinfo {author} {\bibfnamefont {R.~M.}\ \bibnamefont
  {Nandkishore}}, \bibinfo {author} {\bibfnamefont {M.}~\bibnamefont
  {Pretko}},\ and\ \bibinfo {author} {\bibfnamefont {G.}~\bibnamefont
  {Smith}},\ }\bibfield  {title} {\bibinfo {title} {Unitary-projective
  entanglement dynamics},\ }\href {https://doi.org/10.1103/PhysRevB.99.224307}
  {\bibfield  {journal} {\bibinfo  {journal} {Phys. Rev. B}\ }\textbf {\bibinfo
  {volume} {99}},\ \bibinfo {pages} {224307} (\bibinfo {year}
  {2019})}\BibitemShut {NoStop}%
\bibitem [{\citenamefont {Basu}\ \emph {et~al.}(2022)\citenamefont {Basu},
  \citenamefont {Arovas}, \citenamefont {Gopalakrishnan}, \citenamefont
  {Hooley},\ and\ \citenamefont {Oganesyan}}]{PhysRevResearch.4.013018}%
  \BibitemOpen
  \bibfield  {author} {\bibinfo {author} {\bibfnamefont {S.}~\bibnamefont
  {Basu}}, \bibinfo {author} {\bibfnamefont {D.~P.}\ \bibnamefont {Arovas}},
  \bibinfo {author} {\bibfnamefont {S.}~\bibnamefont {Gopalakrishnan}},
  \bibinfo {author} {\bibfnamefont {C.~A.}\ \bibnamefont {Hooley}},\ and\
  \bibinfo {author} {\bibfnamefont {V.}~\bibnamefont {Oganesyan}},\ }\bibfield
  {title} {\bibinfo {title} {Fisher zeros and persistent temporal oscillations
  in nonunitary quantum circuits},\ }\href
  {https://doi.org/10.1103/PhysRevResearch.4.013018} {\bibfield  {journal}
  {\bibinfo  {journal} {Phys. Rev. Res.}\ }\textbf {\bibinfo {volume} {4}},\
  \bibinfo {pages} {013018} (\bibinfo {year} {2022})}\BibitemShut {NoStop}%
\bibitem [{\citenamefont {Takahashi}\ and\ \citenamefont
  {Umezawa}(1996)}]{takahashi1996thermo}%
  \BibitemOpen
  \bibfield  {author} {\bibinfo {author} {\bibfnamefont {Y.}~\bibnamefont
  {Takahashi}}\ and\ \bibinfo {author} {\bibfnamefont {H.}~\bibnamefont
  {Umezawa}},\ }\bibfield  {title} {\bibinfo {title} {Thermo field dynamics},\
  }\href@noop {} {\bibfield  {journal} {\bibinfo  {journal} {International
  journal of modern Physics B}\ }\textbf {\bibinfo {volume} {10}},\ \bibinfo
  {pages} {1755} (\bibinfo {year} {1996})}\BibitemShut {NoStop}%
\bibitem [{\citenamefont {Maldacena}(2003)}]{Maldacena:2003aa}%
  \BibitemOpen
  \bibfield  {author} {\bibinfo {author} {\bibfnamefont {J.}~\bibnamefont
  {Maldacena}},\ }\bibfield  {title} {\bibinfo {title} {Eternal black holes in
  anti-de sitter},\ }\href {https://doi.org/10.1088/1126-6708/2003/04/021}
  {\bibfield  {journal} {\bibinfo  {journal} {Journal of High Energy Physics}\
  }\textbf {\bibinfo {volume} {2003}},\ \bibinfo {pages} {021} (\bibinfo {year}
  {2003})}\BibitemShut {NoStop}%
\bibitem [{\citenamefont {Nivedita}\ \emph {et~al.}(2020)\citenamefont
  {Nivedita}, \citenamefont {Shackleton},\ and\ \citenamefont
  {Sachdev}}]{NiveditaPRE2020}%
  \BibitemOpen
  \bibfield  {author} {\bibinfo {author} {\bibnamefont {Nivedita}}, \bibinfo
  {author} {\bibfnamefont {H.}~\bibnamefont {Shackleton}},\ and\ \bibinfo
  {author} {\bibfnamefont {S.}~\bibnamefont {Sachdev}},\ }\bibfield  {title}
  {\bibinfo {title} {Spectral form factors of clean and random quantum ising
  chains},\ }\href {https://doi.org/10.1103/PhysRevE.101.042136} {\bibfield
  {journal} {\bibinfo  {journal} {Phys. Rev. E}\ }\textbf {\bibinfo {volume}
  {101}},\ \bibinfo {pages} {042136} (\bibinfo {year} {2020})}\BibitemShut
  {NoStop}%
\bibitem [{\citenamefont {Onsager}(1944)}]{Onsager1944}%
  \BibitemOpen
  \bibfield  {author} {\bibinfo {author} {\bibfnamefont {L.}~\bibnamefont
  {Onsager}},\ }\bibfield  {title} {\bibinfo {title} {Crystal statistics. i. a
  two-dimensional model with an order-disorder transition},\ }\href
  {https://doi.org/10.1103/PhysRev.65.117} {\bibfield  {journal} {\bibinfo
  {journal} {Phys. Rev.}\ }\textbf {\bibinfo {volume} {65}},\ \bibinfo {pages}
  {117} (\bibinfo {year} {1944})}\BibitemShut {NoStop}%
\bibitem [{\citenamefont {Kaufman}(1949)}]{Kaufman1949}%
  \BibitemOpen
  \bibfield  {author} {\bibinfo {author} {\bibfnamefont {B.}~\bibnamefont
  {Kaufman}},\ }\bibfield  {title} {\bibinfo {title} {Crystal statistics. ii.
  partition function evaluated by spinor analysis},\ }\href
  {https://doi.org/10.1103/PhysRev.76.1232} {\bibfield  {journal} {\bibinfo
  {journal} {Phys. Rev.}\ }\textbf {\bibinfo {volume} {76}},\ \bibinfo {pages}
  {1232} (\bibinfo {year} {1949})}\BibitemShut {NoStop}%
\bibitem [{\citenamefont {Suzuki}(1976)}]{Suzuki1976}%
  \BibitemOpen
  \bibfield  {author} {\bibinfo {author} {\bibfnamefont {M.}~\bibnamefont
  {Suzuki}},\ }\bibfield  {title} {\bibinfo {title} {Relationship between
  d-dimensional quantal spin systems and (d$+$1)-dimensional ising systems:
  Equivalence, critical exponents and systematic approximants of the partition
  function and spin correlations},\ }\href
  {https://doi.org/10.1143/ptp.56.1454} {\bibfield  {journal} {\bibinfo
  {journal} {Progress of Theoretical Physics}\ }\textbf {\bibinfo {volume}
  {56}},\ \bibinfo {pages} {1454} (\bibinfo {year} {1976})}\BibitemShut
  {NoStop}%
\bibitem [{\citenamefont {Seiberg}\ and\ \citenamefont
  {Shao}(2024)}]{Seiberg2024KW}%
  \BibitemOpen
  \bibfield  {author} {\bibinfo {author} {\bibfnamefont {N.}~\bibnamefont
  {Seiberg}}\ and\ \bibinfo {author} {\bibfnamefont {S.-H.}\ \bibnamefont
  {Shao}},\ }\bibfield  {title} {\bibinfo {title} {{Majorana chain and Ising
  model - (non-invertible) translations, anomalies, and emanant symmetries}},\
  }\href {https://doi.org/10.21468/SciPostPhys.16.3.064} {\bibfield  {journal}
  {\bibinfo  {journal} {SciPost Phys.}\ }\textbf {\bibinfo {volume} {16}},\
  \bibinfo {pages} {064} (\bibinfo {year} {2024})}\BibitemShut {NoStop}%
\bibitem [{\citenamefont {Senthil}(2024)}]{Senthil2024}%
  \BibitemOpen
  \bibfield  {author} {\bibinfo {author} {\bibfnamefont {T.}~\bibnamefont
  {Senthil}},\ }\href@noop {} {\bibinfo {title} {Symmetries without an inverse:
  An illustration through the 1+1-d ising model}} (\bibinfo {year} {2024}),\
  \Eprint {https://arxiv.org/abs/Journal Club for Condensed Matter Physics:
  10.36471/JCCM-February-2024-03} {Journal Club for Condensed Matter Physics:
  10.36471/JCCM-February-2024-03} \BibitemShut {NoStop}%
\bibitem [{sup()}]{supp}%
  \BibitemOpen
  \href@noop {} {}\bibinfo {note} {See the Supplemental Material at
  URL-will-be-inserted-by-publisher for (i) Discussion on the Suzuki solution,
  (ii) structure of Fisher zeros, (iii) comparing the Suzuki solution and CFT
  at the quantum critical point, and (iv) self-similarity of $S$.}\BibitemShut
  {Stop}%
\bibitem [{\citenamefont {Or\'us}(2014)}]{ORUS2014117}%
  \BibitemOpen
  \bibfield  {author} {\bibinfo {author} {\bibfnamefont {R.}~\bibnamefont
  {Or\'us}},\ }\bibfield  {title} {\bibinfo {title} {A practical introduction
  to tensor networks: Matrix product states and projected entangled pair
  states},\ }\href {https://doi.org/https://doi.org/10.1016/j.aop.2014.06.013}
  {\bibfield  {journal} {\bibinfo  {journal} {Annals of Physics}\ }\textbf
  {\bibinfo {volume} {349}},\ \bibinfo {pages} {117} (\bibinfo {year}
  {2014})}\BibitemShut {NoStop}%
\bibitem [{\citenamefont {Cirac}\ \emph {et~al.}(2021)\citenamefont {Cirac},
  \citenamefont {P\'erez-Garc\'{\i}a}, \citenamefont {Schuch},\ and\
  \citenamefont {Verstraete}}]{TNreview1}%
  \BibitemOpen
  \bibfield  {author} {\bibinfo {author} {\bibfnamefont {J.~I.}\ \bibnamefont
  {Cirac}}, \bibinfo {author} {\bibfnamefont {D.}~\bibnamefont
  {P\'erez-Garc\'{\i}a}}, \bibinfo {author} {\bibfnamefont {N.}~\bibnamefont
  {Schuch}},\ and\ \bibinfo {author} {\bibfnamefont {F.}~\bibnamefont
  {Verstraete}},\ }\bibfield  {title} {\bibinfo {title} {Matrix product states
  and projected entangled pair states: Concepts, symmetries, theorems},\ }\href
  {https://doi.org/10.1103/RevModPhys.93.045003} {\bibfield  {journal}
  {\bibinfo  {journal} {Rev. Mod. Phys.}\ }\textbf {\bibinfo {volume} {93}},\
  \bibinfo {pages} {045003} (\bibinfo {year} {2021})}\BibitemShut {NoStop}%
\bibitem [{\citenamefont {Meurice}\ \emph {et~al.}(2022)\citenamefont
  {Meurice}, \citenamefont {Sakai},\ and\ \citenamefont
  {Unmuth-Yockey}}]{TNreview2}%
  \BibitemOpen
  \bibfield  {author} {\bibinfo {author} {\bibfnamefont {Y.}~\bibnamefont
  {Meurice}}, \bibinfo {author} {\bibfnamefont {R.}~\bibnamefont {Sakai}},\
  and\ \bibinfo {author} {\bibfnamefont {J.}~\bibnamefont {Unmuth-Yockey}},\
  }\bibfield  {title} {\bibinfo {title} {Tensor lattice field theory for
  renormalization and quantum computing},\ }\href
  {https://doi.org/10.1103/RevModPhys.94.025005} {\bibfield  {journal}
  {\bibinfo  {journal} {Rev. Mod. Phys.}\ }\textbf {\bibinfo {volume} {94}},\
  \bibinfo {pages} {025005} (\bibinfo {year} {2022})}\BibitemShut {NoStop}%
\bibitem [{\citenamefont {Denbleyker}\ \emph {et~al.}(2014)\citenamefont
  {Denbleyker}, \citenamefont {Liu}, \citenamefont {Meurice}, \citenamefont
  {Qin}, \citenamefont {Xiang}, \citenamefont {Xie}, \citenamefont {Yu},\ and\
  \citenamefont {Zou}}]{Zou2014PRD}%
  \BibitemOpen
  \bibfield  {author} {\bibinfo {author} {\bibfnamefont {A.}~\bibnamefont
  {Denbleyker}}, \bibinfo {author} {\bibfnamefont {Y.}~\bibnamefont {Liu}},
  \bibinfo {author} {\bibfnamefont {Y.}~\bibnamefont {Meurice}}, \bibinfo
  {author} {\bibfnamefont {M.~P.}\ \bibnamefont {Qin}}, \bibinfo {author}
  {\bibfnamefont {T.}~\bibnamefont {Xiang}}, \bibinfo {author} {\bibfnamefont
  {Z.~Y.}\ \bibnamefont {Xie}}, \bibinfo {author} {\bibfnamefont {J.~F.}\
  \bibnamefont {Yu}},\ and\ \bibinfo {author} {\bibfnamefont {H.}~\bibnamefont
  {Zou}},\ }\bibfield  {title} {\bibinfo {title} {Controlling sign problems in
  spin models using tensor renormalization},\ }\href
  {https://doi.org/10.1103/PhysRevD.89.016008} {\bibfield  {journal} {\bibinfo
  {journal} {Phys. Rev. D}\ }\textbf {\bibinfo {volume} {89}},\ \bibinfo
  {pages} {016008} (\bibinfo {year} {2014})}\BibitemShut {NoStop}%
\bibitem [{\citenamefont {Xie}\ \emph {et~al.}(2012)\citenamefont {Xie},
  \citenamefont {Chen}, \citenamefont {Qin}, \citenamefont {Zhu}, \citenamefont
  {Yang},\ and\ \citenamefont {Xiang}}]{XieHOTRG}%
  \BibitemOpen
  \bibfield  {author} {\bibinfo {author} {\bibfnamefont {Z.~Y.}\ \bibnamefont
  {Xie}}, \bibinfo {author} {\bibfnamefont {J.}~\bibnamefont {Chen}}, \bibinfo
  {author} {\bibfnamefont {M.~P.}\ \bibnamefont {Qin}}, \bibinfo {author}
  {\bibfnamefont {J.~W.}\ \bibnamefont {Zhu}}, \bibinfo {author} {\bibfnamefont
  {L.~P.}\ \bibnamefont {Yang}},\ and\ \bibinfo {author} {\bibfnamefont
  {T.}~\bibnamefont {Xiang}},\ }\bibfield  {title} {\bibinfo {title}
  {Coarse-graining renormalization by higher-order singular value
  decomposition},\ }\href {https://doi.org/10.1103/PhysRevB.86.045139}
  {\bibfield  {journal} {\bibinfo  {journal} {Phys. Rev. B}\ }\textbf {\bibinfo
  {volume} {86}},\ \bibinfo {pages} {045139} (\bibinfo {year}
  {2012})}\BibitemShut {NoStop}%
\bibitem [{\citenamefont {Denbleyker}\ \emph {et~al.}(2010)\citenamefont
  {Denbleyker}, \citenamefont {Du}, \citenamefont {Liu}, \citenamefont
  {Meurice},\ and\ \citenamefont {Zou}}]{RGflow2010}%
  \BibitemOpen
  \bibfield  {author} {\bibinfo {author} {\bibfnamefont {A.}~\bibnamefont
  {Denbleyker}}, \bibinfo {author} {\bibfnamefont {D.}~\bibnamefont {Du}},
  \bibinfo {author} {\bibfnamefont {Y.}~\bibnamefont {Liu}}, \bibinfo {author}
  {\bibfnamefont {Y.}~\bibnamefont {Meurice}},\ and\ \bibinfo {author}
  {\bibfnamefont {H.}~\bibnamefont {Zou}},\ }\bibfield  {title} {\bibinfo
  {title} {Fisher's zeros as the boundary of renormalization group flows in
  complex coupling spaces},\ }\href
  {https://doi.org/10.1103/PhysRevLett.104.251601} {\bibfield  {journal}
  {\bibinfo  {journal} {Phys. Rev. Lett.}\ }\textbf {\bibinfo {volume} {104}},\
  \bibinfo {pages} {251601} (\bibinfo {year} {2010})}\BibitemShut {NoStop}%
\bibitem [{\citenamefont {Meurice}\ and\ \citenamefont
  {Zou}(2011)}]{Zou2011PRD}%
  \BibitemOpen
  \bibfield  {author} {\bibinfo {author} {\bibfnamefont {Y.}~\bibnamefont
  {Meurice}}\ and\ \bibinfo {author} {\bibfnamefont {H.}~\bibnamefont {Zou}},\
  }\bibfield  {title} {\bibinfo {title} {Complex renormalization group flows
  for 2d nonlinear $o(n)$ sigma models},\ }\href
  {https://doi.org/10.1103/PhysRevD.83.056009} {\bibfield  {journal} {\bibinfo
  {journal} {Phys. Rev. D}\ }\textbf {\bibinfo {volume} {83}},\ \bibinfo
  {pages} {056009} (\bibinfo {year} {2011})}\BibitemShut {NoStop}%
\bibitem [{\citenamefont {Liu}\ \emph {et~al.}(2024)\citenamefont {Liu},
  \citenamefont {Yin},\ and\ \citenamefont {Chen}}]{Yin2024prb}%
  \BibitemOpen
  \bibfield  {author} {\bibinfo {author} {\bibfnamefont {J.}~\bibnamefont
  {Liu}}, \bibinfo {author} {\bibfnamefont {S.}~\bibnamefont {Yin}},\ and\
  \bibinfo {author} {\bibfnamefont {L.}~\bibnamefont {Chen}},\ }\bibfield
  {title} {\bibinfo {title} {Imaginary-temperature zeros for quantum phase
  transitions},\ }\href {https://doi.org/10.1103/PhysRevB.110.134313}
  {\bibfield  {journal} {\bibinfo  {journal} {Phys. Rev. B}\ }\textbf {\bibinfo
  {volume} {110}},\ \bibinfo {pages} {134313} (\bibinfo {year}
  {2024})}\BibitemShut {NoStop}%
\bibitem [{\citenamefont {Zhang}(2019)}]{Zhang2019prl}%
  \BibitemOpen
  \bibfield  {author} {\bibinfo {author} {\bibfnamefont {L.}~\bibnamefont
  {Zhang}},\ }\bibfield  {title} {\bibinfo {title} {Universal thermodynamic
  signature of self-dual quantum critical points},\ }\href
  {https://doi.org/10.1103/PhysRevLett.123.230601} {\bibfield  {journal}
  {\bibinfo  {journal} {Phys. Rev. Lett.}\ }\textbf {\bibinfo {volume} {123}},\
  \bibinfo {pages} {230601} (\bibinfo {year} {2019})}\BibitemShut {NoStop}%
\bibitem [{\citenamefont {Zhang}\ and\ \citenamefont
  {Ding}(2023)}]{Zhang2023cpl}%
  \BibitemOpen
  \bibfield  {author} {\bibinfo {author} {\bibfnamefont {L.}~\bibnamefont
  {Zhang}}\ and\ \bibinfo {author} {\bibfnamefont {C.}~\bibnamefont {Ding}},\
  }\bibfield  {title} {\bibinfo {title} {Finite-size scaling theory at a
  self-dual quantum critical point},\ }\href
  {https://doi.org/10.1088/0256-307X/40/1/010501} {\bibfield  {journal}
  {\bibinfo  {journal} {Chinese Physics Letters}\ }\textbf {\bibinfo {volume}
  {40}},\ \bibinfo {pages} {010501} (\bibinfo {year} {2023})}\BibitemShut
  {NoStop}%
\bibitem [{\citenamefont {Cotler}\ \emph {et~al.}(2017)\citenamefont {Cotler},
  \citenamefont {Gur-Ari}, \citenamefont {Hanada}, \citenamefont {Polchinski},
  \citenamefont {Saad}, \citenamefont {Shenker}, \citenamefont {Stanford},
  \citenamefont {Streicher},\ and\ \citenamefont {Tezuka}}]{cotler2017black}%
  \BibitemOpen
  \bibfield  {author} {\bibinfo {author} {\bibfnamefont {J.~S.}\ \bibnamefont
  {Cotler}}, \bibinfo {author} {\bibfnamefont {G.}~\bibnamefont {Gur-Ari}},
  \bibinfo {author} {\bibfnamefont {M.}~\bibnamefont {Hanada}}, \bibinfo
  {author} {\bibfnamefont {J.}~\bibnamefont {Polchinski}}, \bibinfo {author}
  {\bibfnamefont {P.}~\bibnamefont {Saad}}, \bibinfo {author} {\bibfnamefont
  {S.~H.}\ \bibnamefont {Shenker}}, \bibinfo {author} {\bibfnamefont
  {D.}~\bibnamefont {Stanford}}, \bibinfo {author} {\bibfnamefont
  {A.}~\bibnamefont {Streicher}},\ and\ \bibinfo {author} {\bibfnamefont
  {M.}~\bibnamefont {Tezuka}},\ }\bibfield  {title} {\bibinfo {title} {Black
  holes and random matrices},\ }\href@noop {} {\bibfield  {journal} {\bibinfo
  {journal} {Journal of High Energy Physics}\ }\textbf {\bibinfo {volume}
  {2017}},\ \bibinfo {pages} {1} (\bibinfo {year} {2017})}\BibitemShut
  {NoStop}%
\bibitem [{\citenamefont {del Campo}\ \emph {et~al.}(2017)\citenamefont {del
  Campo}, \citenamefont {Molina-Vilaplana},\ and\ \citenamefont
  {Sonner}}]{PhysRevD.95.126008}%
  \BibitemOpen
  \bibfield  {author} {\bibinfo {author} {\bibfnamefont {A.}~\bibnamefont {del
  Campo}}, \bibinfo {author} {\bibfnamefont {J.}~\bibnamefont
  {Molina-Vilaplana}},\ and\ \bibinfo {author} {\bibfnamefont {J.}~\bibnamefont
  {Sonner}},\ }\bibfield  {title} {\bibinfo {title} {Scrambling the spectral
  form factor: Unitarity constraints and exact results},\ }\href
  {https://doi.org/10.1103/PhysRevD.95.126008} {\bibfield  {journal} {\bibinfo
  {journal} {Phys. Rev. D}\ }\textbf {\bibinfo {volume} {95}},\ \bibinfo
  {pages} {126008} (\bibinfo {year} {2017})}\BibitemShut {NoStop}%
\bibitem [{\citenamefont {Granet}\ \emph {et~al.}(2023)\citenamefont {Granet},
  \citenamefont {Zhang},\ and\ \citenamefont
  {Dreyer}}]{PhysRevLett.130.230401}%
  \BibitemOpen
  \bibfield  {author} {\bibinfo {author} {\bibfnamefont {E.}~\bibnamefont
  {Granet}}, \bibinfo {author} {\bibfnamefont {C.}~\bibnamefont {Zhang}},\ and\
  \bibinfo {author} {\bibfnamefont {H.}~\bibnamefont {Dreyer}},\ }\bibfield
  {title} {\bibinfo {title} {Volume-law to area-law entanglement transition in
  a nonunitary periodic gaussian circuit},\ }\href
  {https://doi.org/10.1103/PhysRevLett.130.230401} {\bibfield  {journal}
  {\bibinfo  {journal} {Phys. Rev. Lett.}\ }\textbf {\bibinfo {volume} {130}},\
  \bibinfo {pages} {230401} (\bibinfo {year} {2023})}\BibitemShut {NoStop}%
\bibitem [{\citenamefont {Ravindranath}\ and\ \citenamefont
  {Chen}(2023)}]{PhysRevLett.130.230402}%
  \BibitemOpen
  \bibfield  {author} {\bibinfo {author} {\bibfnamefont {V.}~\bibnamefont
  {Ravindranath}}\ and\ \bibinfo {author} {\bibfnamefont {X.}~\bibnamefont
  {Chen}},\ }\bibfield  {title} {\bibinfo {title} {Robust oscillations and edge
  modes in nonunitary floquet systems},\ }\href
  {https://doi.org/10.1103/PhysRevLett.130.230402} {\bibfield  {journal}
  {\bibinfo  {journal} {Phys. Rev. Lett.}\ }\textbf {\bibinfo {volume} {130}},\
  \bibinfo {pages} {230402} (\bibinfo {year} {2023})}\BibitemShut {NoStop}%
\bibitem [{\citenamefont {Su}\ \emph {et~al.}(2024)\citenamefont {Su},
  \citenamefont {Clerk},\ and\ \citenamefont
  {Martin}}]{PhysRevResearch.6.013131}%
  \BibitemOpen
  \bibfield  {author} {\bibinfo {author} {\bibfnamefont {L.}~\bibnamefont
  {Su}}, \bibinfo {author} {\bibfnamefont {A.}~\bibnamefont {Clerk}},\ and\
  \bibinfo {author} {\bibfnamefont {I.}~\bibnamefont {Martin}},\ }\bibfield
  {title} {\bibinfo {title} {Dynamics and phases of nonunitary floquet
  transverse-field ising model},\ }\href
  {https://doi.org/10.1103/PhysRevResearch.6.013131} {\bibfield  {journal}
  {\bibinfo  {journal} {Phys. Rev. Res.}\ }\textbf {\bibinfo {volume} {6}},\
  \bibinfo {pages} {013131} (\bibinfo {year} {2024})}\BibitemShut {NoStop}%
\bibitem [{\citenamefont {Bender}\ and\ \citenamefont
  {Boettcher}(1998)}]{nonHermitian1}%
  \BibitemOpen
  \bibfield  {author} {\bibinfo {author} {\bibfnamefont {C.~M.}\ \bibnamefont
  {Bender}}\ and\ \bibinfo {author} {\bibfnamefont {S.}~\bibnamefont
  {Boettcher}},\ }\bibfield  {title} {\bibinfo {title} {Real spectra in
  non-hermitian hamiltonians having pt symmetry},\ }\href
  {https://doi.org/10.1103/PhysRevLett.80.5243} {\bibfield  {journal} {\bibinfo
   {journal} {Phys. Rev. Lett.}\ }\textbf {\bibinfo {volume} {80}},\ \bibinfo
  {pages} {5243} (\bibinfo {year} {1998})}\BibitemShut {NoStop}%
\bibitem [{\citenamefont {Peng}\ \emph {et~al.}(2015)\citenamefont {Peng},
  \citenamefont {Zhou}, \citenamefont {Wei}, \citenamefont {Cui}, \citenamefont
  {Du},\ and\ \citenamefont {Liu}}]{zero_exp1}%
  \BibitemOpen
  \bibfield  {author} {\bibinfo {author} {\bibfnamefont {X.}~\bibnamefont
  {Peng}}, \bibinfo {author} {\bibfnamefont {H.}~\bibnamefont {Zhou}}, \bibinfo
  {author} {\bibfnamefont {B.-B.}\ \bibnamefont {Wei}}, \bibinfo {author}
  {\bibfnamefont {J.}~\bibnamefont {Cui}}, \bibinfo {author} {\bibfnamefont
  {J.}~\bibnamefont {Du}},\ and\ \bibinfo {author} {\bibfnamefont {R.-B.}\
  \bibnamefont {Liu}},\ }\bibfield  {title} {\bibinfo {title} {Experimental
  observation of lee-yang zeros},\ }\href
  {https://doi.org/10.1103/PhysRevLett.114.010601} {\bibfield  {journal}
  {\bibinfo  {journal} {Phys. Rev. Lett.}\ }\textbf {\bibinfo {volume} {114}},\
  \bibinfo {pages} {010601} (\bibinfo {year} {2015})}\BibitemShut {NoStop}%
\bibitem [{\citenamefont {Jurcevic}\ \emph {et~al.}(2017)\citenamefont
  {Jurcevic}, \citenamefont {Shen}, \citenamefont {Hauke}, \citenamefont
  {Maier}, \citenamefont {Brydges}, \citenamefont {Hempel}, \citenamefont
  {Lanyon}, \citenamefont {Heyl}, \citenamefont {Blatt},\ and\ \citenamefont
  {Roos}}]{DQPT_exp}%
  \BibitemOpen
  \bibfield  {author} {\bibinfo {author} {\bibfnamefont {P.}~\bibnamefont
  {Jurcevic}}, \bibinfo {author} {\bibfnamefont {H.}~\bibnamefont {Shen}},
  \bibinfo {author} {\bibfnamefont {P.}~\bibnamefont {Hauke}}, \bibinfo
  {author} {\bibfnamefont {C.}~\bibnamefont {Maier}}, \bibinfo {author}
  {\bibfnamefont {T.}~\bibnamefont {Brydges}}, \bibinfo {author} {\bibfnamefont
  {C.}~\bibnamefont {Hempel}}, \bibinfo {author} {\bibfnamefont {B.~P.}\
  \bibnamefont {Lanyon}}, \bibinfo {author} {\bibfnamefont {M.}~\bibnamefont
  {Heyl}}, \bibinfo {author} {\bibfnamefont {R.}~\bibnamefont {Blatt}},\ and\
  \bibinfo {author} {\bibfnamefont {C.~F.}\ \bibnamefont {Roos}},\ }\bibfield
  {title} {\bibinfo {title} {Direct observation of dynamical quantum phase
  transitions in an interacting many-body system},\ }\href
  {https://doi.org/10.1103/PhysRevLett.119.080501} {\bibfield  {journal}
  {\bibinfo  {journal} {Phys. Rev. Lett.}\ }\textbf {\bibinfo {volume} {119}},\
  \bibinfo {pages} {080501} (\bibinfo {year} {2017})}\BibitemShut {NoStop}%
\bibitem [{\citenamefont {Zhu}\ \emph {et~al.}(2020)\citenamefont {Zhu},
  \citenamefont {Johri}, \citenamefont {Linke}, \citenamefont {Landsman},
  \citenamefont {Huerta~Alderete}, \citenamefont {Nguyen}, \citenamefont
  {Matsuura}, \citenamefont {Hsieh},\ and\ \citenamefont
  {Monroe}}]{zhu2020generation}%
  \BibitemOpen
  \bibfield  {author} {\bibinfo {author} {\bibfnamefont {D.}~\bibnamefont
  {Zhu}}, \bibinfo {author} {\bibfnamefont {S.}~\bibnamefont {Johri}}, \bibinfo
  {author} {\bibfnamefont {N.~M.}\ \bibnamefont {Linke}}, \bibinfo {author}
  {\bibfnamefont {K.}~\bibnamefont {Landsman}}, \bibinfo {author}
  {\bibfnamefont {C.}~\bibnamefont {Huerta~Alderete}}, \bibinfo {author}
  {\bibfnamefont {N.~H.}\ \bibnamefont {Nguyen}}, \bibinfo {author}
  {\bibfnamefont {A.}~\bibnamefont {Matsuura}}, \bibinfo {author}
  {\bibfnamefont {T.}~\bibnamefont {Hsieh}},\ and\ \bibinfo {author}
  {\bibfnamefont {C.}~\bibnamefont {Monroe}},\ }\bibfield  {title} {\bibinfo
  {title} {Generation of thermofield double states and critical ground states
  with a quantum computer},\ }\href@noop {} {\bibfield  {journal} {\bibinfo
  {journal} {Proceedings of the National Academy of Sciences}\ }\textbf
  {\bibinfo {volume} {117}},\ \bibinfo {pages} {25402} (\bibinfo {year}
  {2020})}\BibitemShut {NoStop}%
\bibitem [{\citenamefont {Wei}\ \emph {et~al.}(2014)\citenamefont {Wei},
  \citenamefont {Chen}, \citenamefont {Po},\ and\ \citenamefont
  {Liu}}]{Wei2014}%
  \BibitemOpen
  \bibfield  {author} {\bibinfo {author} {\bibfnamefont {B.-B.}\ \bibnamefont
  {Wei}}, \bibinfo {author} {\bibfnamefont {S.-W.}\ \bibnamefont {Chen}},
  \bibinfo {author} {\bibfnamefont {H.-C.}\ \bibnamefont {Po}},\ and\ \bibinfo
  {author} {\bibfnamefont {R.-B.}\ \bibnamefont {Liu}},\ }\bibfield  {title}
  {\bibinfo {title} {Phase transitions in the complex plane of physical
  parameters},\ }\bibfield  {journal} {\bibinfo  {journal} {Scientific
  Reports}\ }\textbf {\bibinfo {volume} {4}},\ \href
  {https://doi.org/10.1038/srep05202} {10.1038/srep05202} (\bibinfo {year}
  {2014})\BibitemShut {NoStop}%
\bibitem [{\citenamefont {Chen}\ \emph {et~al.}(2013)\citenamefont {Chen},
  \citenamefont {Gu}, \citenamefont {Liu},\ and\ \citenamefont
  {Wen}}]{Wen2013}%
  \BibitemOpen
  \bibfield  {author} {\bibinfo {author} {\bibfnamefont {X.}~\bibnamefont
  {Chen}}, \bibinfo {author} {\bibfnamefont {Z.-C.}\ \bibnamefont {Gu}},
  \bibinfo {author} {\bibfnamefont {Z.-X.}\ \bibnamefont {Liu}},\ and\ \bibinfo
  {author} {\bibfnamefont {X.-G.}\ \bibnamefont {Wen}},\ }\bibfield  {title}
  {\bibinfo {title} {Symmetry protected topological orders and the group
  cohomology of their symmetry group},\ }\href
  {https://doi.org/10.1103/PhysRevB.87.155114} {\bibfield  {journal} {\bibinfo
  {journal} {Phys. Rev. B}\ }\textbf {\bibinfo {volume} {87}},\ \bibinfo
  {pages} {155114} (\bibinfo {year} {2013})}\BibitemShut {NoStop}%
\bibitem [{\citenamefont {Zou}\ \emph {et~al.}(2019)\citenamefont {Zou},
  \citenamefont {Zhao}, \citenamefont {Guan},\ and\ \citenamefont
  {Liu}}]{Zou2019PRL}%
  \BibitemOpen
  \bibfield  {author} {\bibinfo {author} {\bibfnamefont {H.}~\bibnamefont
  {Zou}}, \bibinfo {author} {\bibfnamefont {E.}~\bibnamefont {Zhao}}, \bibinfo
  {author} {\bibfnamefont {X.-W.}\ \bibnamefont {Guan}},\ and\ \bibinfo
  {author} {\bibfnamefont {W.~V.}\ \bibnamefont {Liu}},\ }\bibfield  {title}
  {\bibinfo {title} {Exactly solvable points and symmetry protected topological
  phases of quantum spins on a zig-zag lattice},\ }\href
  {https://doi.org/10.1103/PhysRevLett.122.180401} {\bibfield  {journal}
  {\bibinfo  {journal} {Phys. Rev. Lett.}\ }\textbf {\bibinfo {volume} {122}},\
  \bibinfo {pages} {180401} (\bibinfo {year} {2019})}\BibitemShut {NoStop}%
\bibitem [{\citenamefont {Zheng}\ \emph {et~al.}(2020)\citenamefont {Zheng},
  \citenamefont {Li},\ and\ \citenamefont {Zou}}]{Zou2020PRB}%
  \BibitemOpen
  \bibfield  {author} {\bibinfo {author} {\bibfnamefont {Q.}~\bibnamefont
  {Zheng}}, \bibinfo {author} {\bibfnamefont {X.}~\bibnamefont {Li}},\ and\
  \bibinfo {author} {\bibfnamefont {H.}~\bibnamefont {Zou}},\ }\bibfield
  {title} {\bibinfo {title} {Symmetry-protected topological phase transitions
  and robust chiral order on a tunable zigzag lattice},\ }\href
  {https://doi.org/10.1103/PhysRevB.101.165131} {\bibfield  {journal} {\bibinfo
   {journal} {Phys. Rev. B}\ }\textbf {\bibinfo {volume} {101}},\ \bibinfo
  {pages} {165131} (\bibinfo {year} {2020})}\BibitemShut {NoStop}%
\bibitem [{\citenamefont {Zhu}\ \emph {et~al.}(2013)\citenamefont {Zhu},
  \citenamefont {Huse},\ and\ \citenamefont {White}}]{Zhu2013PRL}%
  \BibitemOpen
  \bibfield  {author} {\bibinfo {author} {\bibfnamefont {Z.}~\bibnamefont
  {Zhu}}, \bibinfo {author} {\bibfnamefont {D.~A.}\ \bibnamefont {Huse}},\ and\
  \bibinfo {author} {\bibfnamefont {S.~R.}\ \bibnamefont {White}},\ }\bibfield
  {title} {\bibinfo {title} {Unexpected $z$-direction ising antiferromagnetic
  order in a frustrated spin-$1/2$ ${J}_{1}\ensuremath{-}{J}_{2}$ $xy$ model on
  the honeycomb lattice},\ }\href
  {https://doi.org/10.1103/PhysRevLett.111.257201} {\bibfield  {journal}
  {\bibinfo  {journal} {Phys. Rev. Lett.}\ }\textbf {\bibinfo {volume} {111}},\
  \bibinfo {pages} {257201} (\bibinfo {year} {2013})}\BibitemShut {NoStop}%
\bibitem [{\citenamefont {Liu}\ \emph {et~al.}(2022)\citenamefont {Liu},
  \citenamefont {Gong}, \citenamefont {Li}, \citenamefont {Poilblanc},
  \citenamefont {Chen},\ and\ \citenamefont {Gu}}]{Liu2022}%
  \BibitemOpen
  \bibfield  {author} {\bibinfo {author} {\bibfnamefont {W.-Y.}\ \bibnamefont
  {Liu}}, \bibinfo {author} {\bibfnamefont {S.-S.}\ \bibnamefont {Gong}},
  \bibinfo {author} {\bibfnamefont {Y.-B.}\ \bibnamefont {Li}}, \bibinfo
  {author} {\bibfnamefont {D.}~\bibnamefont {Poilblanc}}, \bibinfo {author}
  {\bibfnamefont {W.-Q.}\ \bibnamefont {Chen}},\ and\ \bibinfo {author}
  {\bibfnamefont {Z.-C.}\ \bibnamefont {Gu}},\ }\bibfield  {title} {\bibinfo
  {title} {Gapless quantum spin liquid and global phase diagram of the spin-1/2
  j1-j2 square antiferromagnetic heisenberg model},\ }\href
  {https://doi.org/10.1016/j.scib.2022.03.010} {\bibfield  {journal} {\bibinfo
  {journal} {Science Bulletin}\ }\textbf {\bibinfo {volume} {67}},\ \bibinfo
  {pages} {1034} (\bibinfo {year} {2022})}\BibitemShut {NoStop}%
\bibitem [{\citenamefont {Zou}\ \emph {et~al.}(2024)\citenamefont {Zou},
  \citenamefont {Yang},\ and\ \citenamefont {Ku}}]{Zou2023}%
  \BibitemOpen
  \bibfield  {author} {\bibinfo {author} {\bibfnamefont {H.}~\bibnamefont
  {Zou}}, \bibinfo {author} {\bibfnamefont {F.}~\bibnamefont {Yang}},\ and\
  \bibinfo {author} {\bibfnamefont {W.}~\bibnamefont {Ku}},\ }\bibfield
  {title} {\bibinfo {title} {Nearly degenerate ground states of a checkerboard
  antiferromagnet and their bosonic interpretation},\ }\href
  {http://dx.doi.org/10.1007/s11433-023-2190-5} {\bibfield  {journal} {\bibinfo
   {journal} {Sci. China Phys. Mech. Astron.}\ }\textbf {\bibinfo {volume}
  {67}},\ \bibinfo {pages} {217211} (\bibinfo {year} {2024})}\BibitemShut
  {NoStop}%
\end{thebibliography}

\begin{thebibliography}{12}%
\makeatletter
\providecommand \@ifxundefined [1]{%
 \@ifx{#1\undefined}
}%
\providecommand \@ifnum [1]{%
 \ifnum #1\expandafter \@firstoftwo
 \else \expandafter \@secondoftwo
 \fi
}%
\providecommand \@ifx [1]{%
 \ifx #1\expandafter \@firstoftwo
 \else \expandafter \@secondoftwo
 \fi
}%
\providecommand \natexlab [1]{#1}%
\providecommand \enquote  [1]{``#1''}%
\providecommand \bibnamefont  [1]{#1}%
\providecommand \bibfnamefont [1]{#1}%
\providecommand \citenamefont [1]{#1}%
\providecommand \href@noop [0]{\@secondoftwo}%
\providecommand \href [0]{\begingroup \@sanitize@url \@href}%
\providecommand \@href[1]{\@@startlink{#1}\@@href}%
\providecommand \@@href[1]{\endgroup#1\@@endlink}%
\providecommand \@sanitize@url [0]{\catcode `\\12\catcode `\$12\catcode
  `\&12\catcode `\#12\catcode `\^12\catcode `\_12\catcode `\%12\relax}%
\providecommand \@@startlink[1]{}%
\providecommand \@@endlink[0]{}%
\providecommand \url  [0]{\begingroup\@sanitize@url \@url }%
\providecommand \@url [1]{\endgroup\@href {#1}{\urlprefix }}%
\providecommand \urlprefix  [0]{URL }%
\providecommand \Eprint [0]{\href }%
\providecommand \doibase [0]{https://doi.org/}%
\providecommand \selectlanguage [0]{\@gobble}%
\providecommand \bibinfo  [0]{\@secondoftwo}%
\providecommand \bibfield  [0]{\@secondoftwo}%
\providecommand \translation [1]{[#1]}%
\providecommand \BibitemOpen [0]{}%
\providecommand \bibitemStop [0]{}%
\providecommand \bibitemNoStop [0]{.\EOS\space}%
\providecommand \EOS [0]{\spacefactor3000\relax}%
\providecommand \BibitemShut  [1]{\csname bibitem#1\endcsname}%
\let\auto@bib@innerbib\@empty
\bibitem [{\citenamefont {Suzuki}(1976)}]{Suzuki1976S}%
  \BibitemOpen
  \bibfield  {author} {\bibinfo {author} {\bibfnamefont {M.}~\bibnamefont
  {Suzuki}},\ }\bibfield  {title} {\bibinfo {title} {Relationship between
  d-dimensional quantal spin systems and (d$+$1)-dimensional ising systems:
  Equivalence, critical exponents and systematic approximants of the partition
  function and spin correlations},\ }\href
  {https://doi.org/10.1143/ptp.56.1454} {\bibfield  {journal} {\bibinfo
  {journal} {Progress of Theoretical Physics}\ }\textbf {\bibinfo {volume}
  {56}},\ \bibinfo {pages} {1454} (\bibinfo {year} {1976})}\BibitemShut
  {NoStop}%
\bibitem [{\citenamefont {Kramers}\ and\ \citenamefont
  {Wannier}(1941)}]{KW1941S}%
  \BibitemOpen
  \bibfield  {author} {\bibinfo {author} {\bibfnamefont {H.~A.}\ \bibnamefont
  {Kramers}}\ and\ \bibinfo {author} {\bibfnamefont {G.~H.}\ \bibnamefont
  {Wannier}},\ }\bibfield  {title} {\bibinfo {title} {Statistics of the
  two-dimensional ferromagnet. part i},\ }\href
  {https://doi.org/10.1103/PhysRev.60.252} {\bibfield  {journal} {\bibinfo
  {journal} {Phys. Rev.}\ }\textbf {\bibinfo {volume} {60}},\ \bibinfo {pages}
  {252} (\bibinfo {year} {1941})}\BibitemShut {NoStop}%
\bibitem [{\citenamefont {Liu}\ \emph {et~al.}(2023)\citenamefont {Liu},
  \citenamefont {Lv}, \citenamefont {Yang},\ and\ \citenamefont
  {Zou}}]{liu2023CPLS}%
  \BibitemOpen
  \bibfield  {author} {\bibinfo {author} {\bibfnamefont {Y.}~\bibnamefont
  {Liu}}, \bibinfo {author} {\bibfnamefont {S.}~\bibnamefont {Lv}}, \bibinfo
  {author} {\bibfnamefont {Y.}~\bibnamefont {Yang}},\ and\ \bibinfo {author}
  {\bibfnamefont {H.}~\bibnamefont {Zou}},\ }\bibfield  {title} {\bibinfo
  {title} {Signatures of quantum criticality in the complex inverse temperature
  plane},\ }\href@noop {} {\bibfield  {journal} {\bibinfo  {journal} {Chinese
  Physics Letters}\ }\textbf {\bibinfo {volume} {40}},\ \bibinfo {pages}
  {050502} (\bibinfo {year} {2023})}\BibitemShut {NoStop}%
\bibitem [{\citenamefont {Xie}\ \emph {et~al.}(2012)\citenamefont {Xie},
  \citenamefont {Chen}, \citenamefont {Qin}, \citenamefont {Zhu}, \citenamefont
  {Yang},\ and\ \citenamefont {Xiang}}]{XieHOTRGS}%
  \BibitemOpen
  \bibfield  {author} {\bibinfo {author} {\bibfnamefont {Z.~Y.}\ \bibnamefont
  {Xie}}, \bibinfo {author} {\bibfnamefont {J.}~\bibnamefont {Chen}}, \bibinfo
  {author} {\bibfnamefont {M.~P.}\ \bibnamefont {Qin}}, \bibinfo {author}
  {\bibfnamefont {J.~W.}\ \bibnamefont {Zhu}}, \bibinfo {author} {\bibfnamefont
  {L.~P.}\ \bibnamefont {Yang}},\ and\ \bibinfo {author} {\bibfnamefont
  {T.}~\bibnamefont {Xiang}},\ }\bibfield  {title} {\bibinfo {title}
  {Coarse-graining renormalization by higher-order singular value
  decomposition},\ }\href {https://doi.org/10.1103/PhysRevB.86.045139}
  {\bibfield  {journal} {\bibinfo  {journal} {Phys. Rev. B}\ }\textbf {\bibinfo
  {volume} {86}},\ \bibinfo {pages} {045139} (\bibinfo {year}
  {2012})}\BibitemShut {NoStop}%
\bibitem [{\citenamefont {Denbleyker}\ \emph {et~al.}(2014)\citenamefont
  {Denbleyker}, \citenamefont {Liu}, \citenamefont {Meurice}, \citenamefont
  {Qin}, \citenamefont {Xiang}, \citenamefont {Xie}, \citenamefont {Yu},\ and\
  \citenamefont {Zou}}]{Zou2014PRDS}%
  \BibitemOpen
  \bibfield  {author} {\bibinfo {author} {\bibfnamefont {A.}~\bibnamefont
  {Denbleyker}}, \bibinfo {author} {\bibfnamefont {Y.}~\bibnamefont {Liu}},
  \bibinfo {author} {\bibfnamefont {Y.}~\bibnamefont {Meurice}}, \bibinfo
  {author} {\bibfnamefont {M.~P.}\ \bibnamefont {Qin}}, \bibinfo {author}
  {\bibfnamefont {T.}~\bibnamefont {Xiang}}, \bibinfo {author} {\bibfnamefont
  {Z.~Y.}\ \bibnamefont {Xie}}, \bibinfo {author} {\bibfnamefont {J.~F.}\
  \bibnamefont {Yu}},\ and\ \bibinfo {author} {\bibfnamefont {H.}~\bibnamefont
  {Zou}},\ }\bibfield  {title} {\bibinfo {title} {Controlling sign problems in
  spin models using tensor renormalization},\ }\href
  {https://doi.org/10.1103/PhysRevD.89.016008} {\bibfield  {journal} {\bibinfo
  {journal} {Phys. Rev. D}\ }\textbf {\bibinfo {volume} {89}},\ \bibinfo
  {pages} {016008} (\bibinfo {year} {2014})}\BibitemShut {NoStop}%
\bibitem [{\citenamefont {Kaufman}(1949)}]{Kaufman1949S}%
  \BibitemOpen
  \bibfield  {author} {\bibinfo {author} {\bibfnamefont {B.}~\bibnamefont
  {Kaufman}},\ }\bibfield  {title} {\bibinfo {title} {Crystal statistics. ii.
  partition function evaluated by spinor analysis},\ }\href
  {https://doi.org/10.1103/PhysRev.76.1232} {\bibfield  {journal} {\bibinfo
  {journal} {Phys. Rev.}\ }\textbf {\bibinfo {volume} {76}},\ \bibinfo {pages}
  {1232} (\bibinfo {year} {1949})}\BibitemShut {NoStop}%
\bibitem [{\citenamefont {Fisher}\ and\ \citenamefont
  {Brittin}(1965)}]{fisher1965statisticalS}%
  \BibitemOpen
  \bibfield  {author} {\bibinfo {author} {\bibfnamefont {M.}~\bibnamefont
  {Fisher}}\ and\ \bibinfo {author} {\bibfnamefont {W.}~\bibnamefont
  {Brittin}},\ }\bibfield  {title} {\bibinfo {title} {Statistical physics, weak
  interactions, field theory},\ }\href@noop {} {\bibfield  {journal} {\bibinfo
  {journal} {Lectures in Theoretical Physics (Boulder: University of Colorado
  Press) vol VIIC}\ } (\bibinfo {year} {1965})}\BibitemShut {NoStop}%
\bibitem [{\citenamefont {Bena}\ \emph {et~al.}(2005)\citenamefont {Bena},
  \citenamefont {Droz},\ and\ \citenamefont {Lipowski}}]{bena2005statisticalS}%
  \BibitemOpen
  \bibfield  {author} {\bibinfo {author} {\bibfnamefont {I.}~\bibnamefont
  {Bena}}, \bibinfo {author} {\bibfnamefont {M.}~\bibnamefont {Droz}},\ and\
  \bibinfo {author} {\bibfnamefont {A.}~\bibnamefont {Lipowski}},\ }\bibfield
  {title} {\bibinfo {title} {Statistical mechanics of equilibrium and
  nonequilibrium phase transitions: the yang--lee formalism},\ }\href@noop {}
  {\bibfield  {journal} {\bibinfo  {journal} {International Journal of Modern
  Physics B}\ }\textbf {\bibinfo {volume} {19}},\ \bibinfo {pages} {4269}
  (\bibinfo {year} {2005})}\BibitemShut {NoStop}%
\bibitem [{\citenamefont {Zhang}(2019)}]{Zhang2019prlS}%
  \BibitemOpen
  \bibfield  {author} {\bibinfo {author} {\bibfnamefont {L.}~\bibnamefont
  {Zhang}},\ }\bibfield  {title} {\bibinfo {title} {Universal thermodynamic
  signature of self-dual quantum critical points},\ }\href
  {https://doi.org/10.1103/PhysRevLett.123.230601} {\bibfield  {journal}
  {\bibinfo  {journal} {Phys. Rev. Lett.}\ }\textbf {\bibinfo {volume} {123}},\
  \bibinfo {pages} {230601} (\bibinfo {year} {2019})}\BibitemShut {NoStop}%
\bibitem [{\citenamefont {Zhang}\ and\ \citenamefont
  {Ding}(2023)}]{Zhang2023cplS}%
  \BibitemOpen
  \bibfield  {author} {\bibinfo {author} {\bibfnamefont {L.}~\bibnamefont
  {Zhang}}\ and\ \bibinfo {author} {\bibfnamefont {C.}~\bibnamefont {Ding}},\
  }\bibfield  {title} {\bibinfo {title} {Finite-size scaling theory at a
  self-dual quantum critical point},\ }\href
  {https://doi.org/10.1088/0256-307X/40/1/010501} {\bibfield  {journal}
  {\bibinfo  {journal} {Chinese Physics Letters}\ }\textbf {\bibinfo {volume}
  {40}},\ \bibinfo {pages} {010501} (\bibinfo {year} {2023})}\BibitemShut
  {NoStop}%
\bibitem [{\citenamefont {Di~Francesco}\ \emph {et~al.}(2011)\citenamefont
  {Di~Francesco}, \citenamefont {Mathieu},\ and\ \citenamefont
  {Senechal}}]{CFTbookS}%
  \BibitemOpen
  \bibfield  {author} {\bibinfo {author} {\bibfnamefont {P.}~\bibnamefont
  {Di~Francesco}}, \bibinfo {author} {\bibfnamefont {P.}~\bibnamefont
  {Mathieu}},\ and\ \bibinfo {author} {\bibfnamefont {D.}~\bibnamefont
  {Senechal}},\ }\href@noop {} {\emph {\bibinfo {title} {Conformal Field
  Theory}}}\ (\bibinfo  {publisher} {Springer},\ \bibinfo {year}
  {2011})\BibitemShut {NoStop}%
\bibitem [{\citenamefont {Nivedita}\ \emph {et~al.}(2020)\citenamefont
  {Nivedita}, \citenamefont {Shackleton},\ and\ \citenamefont
  {Sachdev}}]{NiveditaPRE2020S}%
  \BibitemOpen
  \bibfield  {author} {\bibinfo {author} {\bibnamefont {Nivedita}}, \bibinfo
  {author} {\bibfnamefont {H.}~\bibnamefont {Shackleton}},\ and\ \bibinfo
  {author} {\bibfnamefont {S.}~\bibnamefont {Sachdev}},\ }\bibfield  {title}
  {\bibinfo {title} {Spectral form factors of clean and random quantum ising
  chains},\ }\href {https://doi.org/10.1103/PhysRevE.101.042136} {\bibfield
  {journal} {\bibinfo  {journal} {Phys. Rev. E}\ }\textbf {\bibinfo {volume}
  {101}},\ \bibinfo {pages} {042136} (\bibinfo {year} {2020})}\BibitemShut
  {NoStop}%
\end{thebibliography}
%

\end{document}